# EXTRAGALACTIC RESULTS FROM THE INFRARED SPACE OBSERVATORY

Reinhard Genzel[1][2] and Catherine J. Cesarsky[3][4]

[1] Max-Planck Institut für extraterrestrische Physik (MPE), Garching, FRG

genzel@mpe-garching.mpg.de

[2] Dept. of Physics, University of California, Berkeley, USA

[3] Service d'Astrophysique, DAPNIA/DSM, CEA Saclay, Fr

[4] European Southern Observatory, Garching, FRG

ccesarsk@eso.org

*Proofs to be sent to:*

Prof. Reinhard Genzel

Max-Planck-Institut für extraterrestrische Physik

Postfach 1603

85740 Garching, Germany

Phone: +49-89-3299-3280, Fax: +49-89-3299-3601

# ABSTRACT

More than a decade ago the IRAS satellite opened the realm of external galaxies for studies in the 10 to 100µm band and discovered emission from tens of thousands of normal and active galaxies.With the 1995-1998 mission of the Infrared Space Observatory[1] the next major steps in extragalactic infrared astronomy became possible: detailed imaging, spectroscopy and spectro-photometry of many galaxies detected by IRAS, as well as deep surveys in the mid- and far-IR. The spectroscopic data reveal a wealth of detail about the nature of the energy source(s) and about the physical conditions in galaxies. ISO's surveys for the first time explore the infrared emission of distant, high-redshift galaxies. ISO's main theme in extragalactic astronomy is the role of star formation in the activity and evolution of galaxies.

---

[1] The Infrared Space Observatory (ISO) is a project of the European Space Agency (ESA). The satellite and its instruments were funded by the ESA member states (especially the PI countries: France, Germany, the Netherlands and the United Kingdom) and with the participation of the Japanese Space Agency ISAS and NASA.

# TABLE OF CONTENTS

# 1. INTRODUCTION: FROM IRAS TO ISO

In 1983 the first cryogenic infrared astronomy satellite, IRAS[2], surveyed 96% of the sky in four broad-band filters at 12, 25, 60 and 100µm, to limits ≤1 Jy (Neugebauer et al 1984, see reviews of Beichman 1987, Soifer et al 1987). IRAS detected infrared (IR) emission from about 25,000 galaxies, primarily from spirals, but also from quasars (QSOs) (Neugebauer et al 1986, Sanders et al 1989), Seyfert galaxies (de Grijp et al 1985) and early type galaxies (Knapp et al 1989, 1992). IRAS discovered a new class of galaxies which radiate most of their energy in the infrared (Soifer et al 1984), many of them dusty 'starburst' galaxies[3]. The most luminous of these infrared galaxies ('(ultra-) luminous infrared galaxies': (U)LIRGs or (U)LIGs) have QSO-like bolometric luminosities (LIRGs: $L \geq 10^{11} L_\odot$, ULIRGs: $L \geq 10^{12} L_\odot$, Sanders & Mirabel 1996).

The Infrared Space Observatory (Kessler et al 1996) was the first cryogenic space infrared observatory. ISO was launched in November 1995. It was equipped with a multi-pixel near-/mid-IR camera (ISOCAM, CJ Cesarsky et al 1996,1999), a multi-band mid- and far-IR spectro-photometer (ISOPHOT, Lemke et al 1996,1999), a 2.4 to 45µm spectrometer (SWS, deGraauw et al 1996,1999) and a 43-197µm spectrometer (LWS, Clegg et al 1996,1999). The ISO mission lasted until April 1998, about one year longer than expected (Kessler 1999). The present review is an account of the key extragalactic results of ISO as of December 1999. We also refer the reader to the special Astronomy and Astrophysics issue on early ISO results (volume 315, No.2, 1996), to the proceedings of the 1998 Paris conference 'The Universe as seen by ISO' (Cox & Kessler 1999) and to the recent review by C Cesarsky and M Sauvage (1999).

---

[2] The **I**nfra-**R**ed **A**stronomical **S**atellite was developed and operated by the US National Aeronautics and Space Administration ( NASA), the Netherlands Agency for Aerospace Programs (NIVP) and the United Kingdom Science and Engineering Research Council (SERC).

[3] Following the classical analysis by Rieke et al (1980) of two nearby representatives of this class, M82 and NGC 253, these galaxies are presently going through a very active, but short-lived 'starburst' of duration of a few tens of millions of years or less. For a 'Salpeter' initial mass function (IMF) from 100 to 1 $M_\odot$ a luminosity of $10^{10} L_\odot$ corresponds to a star formation rate of roughly 1 $M_\odot$ yr$^{-1}$ (e.g. Kennicutt 1998). Given its infrared luminosity ($\geq 4 \times 10^{10} L_\odot$) and central gas content ($2.5 \times 10^8 M_\odot$), the relatively small galaxy M82 thus cannot sustain its present star formation rate for much longer than 50 million years, hence the term 'starburst'.

# 2. NORMAL GALAXIES

## 2.1 Infrared emission from Normal Galaxies

### *2.1.1 Mid-infrared emission: unidentified bands and very small dust grains*

ISO studies of our Galaxy have demonstrated that the bulk of the mid-IR emission from the interstellar medium is due to transient heating of small clusters of particles (Boulanger et al 1998), thus confirming a longstanding hypothesis based on ground-based and IRAS data (Sellgren 1984, Beichman 1987). The clusters are stochastically heated by single photons and as a result exhibit large temperature fluctuations. The resulting mid-IR flux is simply proportional to the underlying radiation field intensity. The spectra are surprisingly regular, exhibiting almost invariably a family of features centered at 3.3, 6.2, 7.7, 8.6, 11.3 and 12.7μm (cf. Allamandola et al 1995, D. Cesarsky et al. 1996a,b, Verstraete et al 1996, Mattila et al 1996, 1999, Onaka et al 1996). These 'unidentified infrared bands' (UIBs) are thought to be due to C-C and C-H stretching/bending vibrational bands in aromatic hydrocarbons (Puget & Leger 1989, Duley and Williams 1991, Tielens et al 1999). The actual carrier of the bands is still unknown. One possibility is that it consists of large, carbon-rich ring molecules (e.g. polycyclic aromatic hydrocarbons [PAHs], size ≤a few nm). Alternatively the carrier may be very small, amorphous carbon dust grains that are exposed to moderately intense UV (and possibly visible: Boulade et al 1996, Uchida et al. 1998, Pagani et al. 1999) radiation. The most popular model is the PAH-interpretation but no rigorous identification with specific molecules has yet been established. A key prediction of the PAH model, that the features consist of a superposition of a large number of narrow lines, has not been verified by ISO (Puget 1998).

The second component of the interstellar mid-IR emission manifests itself as a steeply rising continuum longward of 10μm, accompanied by strong fine structure line emission, in particular [NeII] and [NeIII]. This continuum component is characteristic of active star forming regions, such as the Galactic HII region M17 (Verstraete et al 1996, D.Cesarky et al 1996b). Desert et al (1990) attribute this continuum to very small, fluctuating grains (VSGs: size ≤10 nm), and this hypothesis is consistent with the ISOCAM CVF spectral maps obtained around M17 and the photometry of NGC 7023 (Tran 1998, Laureijs et al. 1996).

The transition from stellar emission to interstellar dust emission in galaxies occurs in the mid-IR (3 to 30μm) band and depends on the star forming activity. This is well demonstrated by a detailed ISOCAM study of ~$10^2$ Virgo cluster galaxies (Boselli et al 1997, 1999). There are three main components of dust emission contributing to the mid-IR spectra of galaxies. The first is a UIB dominated, mid-IR spectral energy distribution (SED) up to 13 microns. The SED and the ratios of the different UIB features remain constant in galaxies with a wide range of radiation fields and properties (Helou 1999, Helou et al 1999). The second component, present only in some galaxies or regions of galaxies, is the steeply rising (VSG) continuum longwards of 10 microns discussed above. It is characteristic of intense star forming regions. The third is near radiation equilibrium emission from hot (150 to 1700 K) dust particles. This near-/mid-IR 'bump' is characteristic of dust 'tori' in AGNs (section 3.3.1). There may also be a 3-5µm continuum component that may come from a fluctuating dust component without PAH features (Helou et al 1999).

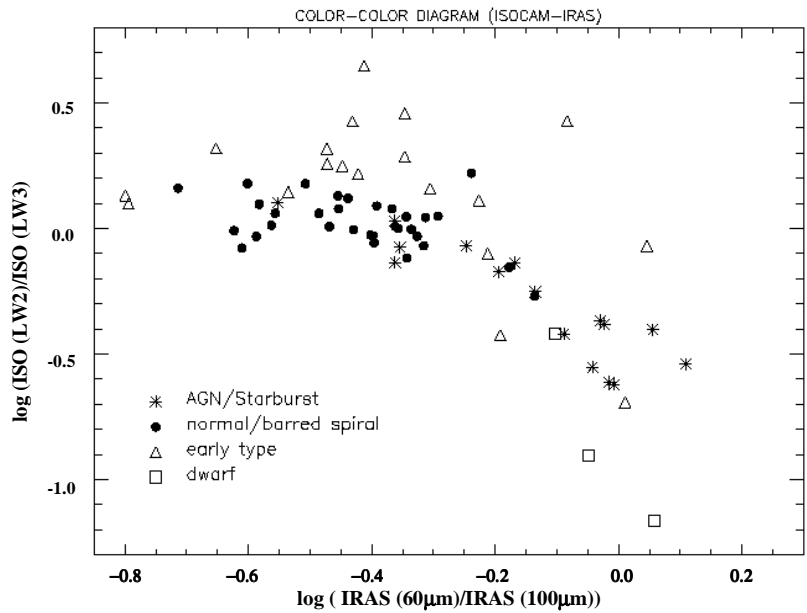

**Fig.1**

UIB emission is a good tracer of normal and moderately active star formation activity in spiral and irregular galaxies (Helou 1999, Vigroux et al 1999). In such systems the luminosity in the ISOCAM LW2 filter (containing the 6.2, 7.7 and 8.6µm UIB features) is well correlated with the longer wavelength mid-IR (LW3 filter: 15µm, Fig.1). Both correlate with the far-IR and the Hα line luminosities (Rouan et al 1996,

Metcalfe et al 1996, Smith 1998, Vigroux et al 1999, Roussel et al 1999a). In normal galaxies with low activity, as in NGC891, NGC 7331, M31 or in parts of the LMC, the mid-IR emission mainly traces the distribution of (molecular + atomic) gas that is bathed in the diffuse radiation field (Mattila et al 1999, Smith 1998, Pagani et al 1999, Contursi et al 1998). Under these conditions excitation of the UIB features other than by UV photons (e.g. visible photons) may play an important role (Pagani et al 1999). At higher luminosities and activity, however, the $\lambda \geq 10\mu m$ continuum (from warm dust in PDRs and HII regions) measured in the LW3 filter increases relative to the UIB emission (Fig.1, Vigroux et al 1999). At radiation fields $\geq 10^5$ times the local interstellar radiation field, such as in the central parsec of the Galaxy (Lutz et al 1996b) or in the M17 HII region (Verstraete et al 1996), in active galactic nuclei (see 3.1), or in metal poor environments (such as the metal poor blue compact dwarf galaxy SBS0335-052, Thuan et al 1999), the UIB emission strength plumets, presumably due to the destruction of its carriers. As a result, the LW3/LW2 ratio is an interesting diagnostic of the radiation environment (Fig.1). This ratio decreases outward from the nuclei in disk galaxies (Dale et al 1999).

*2.1.2 Far-IR emission: cold dust and galactic extinction*

In the far-IR the emitted spectrum results from the radiation equilibrium between absorbed short-wavelength radiation and grey-body emission with a wavelength dependent emissivity ($\epsilon(\lambda) \propto \lambda^{-\beta}$ with $\beta \sim 1.5$ to 2 in the range $\lambda \geq 20\mu m$, Draine & Lee 1984). IRAS observations have established that the $\lambda \leq 100$ μm dust emission in normal spirals comes from ~30 K dust grains with a total gas mass (HI+$H_2$+He) to dust mass ratio of $\geq 10^3$ (e.g. Devereux and Young 1990). This value is about an order of magnitude greater than that in the Milky Way ($M_{gas}/M_{dust}$~170, cf. Haas et al 1998b). The likely source of this discrepancy is that IRAS did not pick up the coldest dust that contains most of the mass (e.g. Xu and Helou 1996).

By extending the wavelength coverage to 200μm ISO has resolved this puzzle. ISOPHOT observations of a number of normal, inactive spirals have uncovered a dust component with typical temperatures ~20 K and a range between 10 and 28 K (Alton et al 1998a, Davies et al 1999, Domingue et al 1999, Haas 1998, Haas et al 1998b, Hippelein et al 1996, Krügel et al 1998, Siebenmorgen et al 1999, Trewhella et al 1997, 1998, Tuffs et al 1996). With increasing activity a second dust component (T~30-40 K) becomes more prominent which also dominates the 60 and 100μm IRAS bands (e.g. Siebenmorgen et al. 1999). The cold dust has a larger radial extent than the stars (Alton et al. 1998a, Davies et al 1999) and may be partially associated with extended HI disk.

First results from the 175μm serendipity survey indicate that this is a general result for normal spirals (Stickel et al 1999). However, Domingue et al (1999) argue from a direct comparison of visual and far-IR data in 3 overlapping galaxies against any dominant part of the dust mass being very much colder than 20 K. Particularly impressive is the case of M31 where Haas et al (1998b) have presented a 175μm map of the entire galaxy (Fig.2). Across most of the disk the far-IR SED is still rising from 100 to 200μm with a temperature of $16\pm2$ K (see also Odenwald et al 1998). The temperature of the cold dust component is consistent with theoretical predictions from the heating/cooling balance of equilibrium dust grains in the diffuse radiation field.

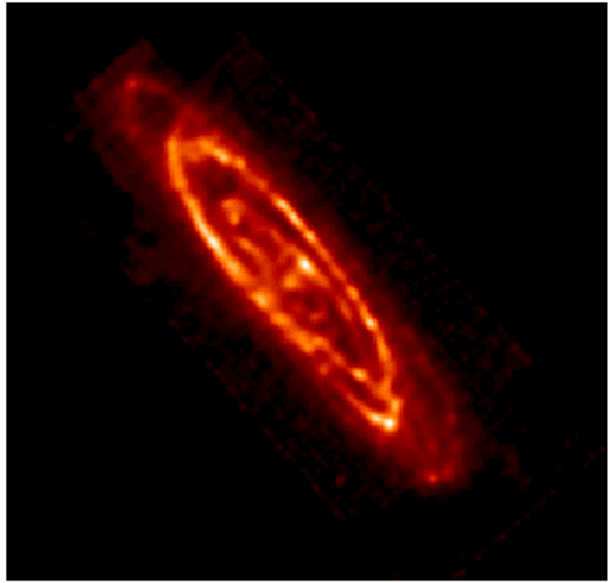

**Fig.2**

The gas to dust ratio inferred for this cold dust component is very close to the Galactic value.

An important consequence of the larger dust masses derived from the ISOPHOT data is that the corresponding dust extinctions in the optical become quite significant. The extinction corrected morphologies of spirals can thus be quite different from the uncorrected ones (Trewhella et al 1997, Haas et al 1998b). For instance, by including extinction M31 turns from an Sb into a ring galaxy (Fig.2) and NGC 6946 from an Sc into an Sb (Trewhella et al. 1998). These findings qualitatively support earlier suggestions that spiral galaxies are affected significantly by dust (Disney et al 1989, Valentijn 1990).

## 2.2 Dark Matter in the Halos and Outer Parts of Disks

Evidence has been mounting for several decades that the outer parts of disk galaxies are dominated by dark matter in the form of a spheroidal, halo distribution (e.g. Trimble 1987, Carr 1994). Of the various possibilities for the nature of this halo, dark baryonic matter in the form of stellar remnants, low mass stars and sub-stellar objects is the most conservative solution, and is still consistent with current Big Bang nucleosynthesis limits (Carr 1994). It is of substantial interest to complement the existing deep optical and near-infrared observations with a search for possible IR halos in edge on galaxies. Such halos could signal the presence of cool, low mass stars or of sub-stellar (≤0.08 $M_\odot$) brown dwarfs.

With ISOCAM Gilmore & Unavane (1998) and Beichman et al (1999) have searched for IR halos in five edge-on galaxies. No off-disk halos are detected in any of them. Gilmore and Unavane (1998) conclude that main sequence, hydrogen burning stars of all masses and metallicities down to the hydrogen burning limit are excluded to dominate the halos. A mixture of low mass stars with a mass function similar to that in the solar neighborhood can also be strongly ruled out in one galaxy (NGC 2915). Young (≤1 Gyr) brown dwarfs are weakly ruled out in UGC 1459. In contrast the EROS/MACHO microlensing data toward the Large Magellanic Cloud favor a mean mass of the deflectors in the halo of our own Galaxy of about 0.5 $M_\odot$ (Alcock et al 1998). A possible way out may be that the lensing objects are not in the Galactic halo but in the LMC (Sahu 1994) or in a debris, dwarf galaxy in between the Galaxy and the LMC (Zhao 1998). The combined ISO, HST and MACHO/EROS results with the additional assumption that the composition of halos is similar in the different galaxies then suggests that with the exception of brown-dwarfs between ≤0.01 and 0.08 $M_\odot$, the entire range of possible stellar and sub-stellar entities is excluded or highly unlikely (Gilmore 1998).

Another possible form of baryonic dark matter in the outer disks of spirals may be cold molecular gas (e.g. Pfenniger et al 1994). There is not much direct evidence for such a component from CO mm-spectroscopy but in the outer, low metallicity regions CO may have low abundance. Evidence for cold dust more extensive than the stellar distributions comes from 200μm ISOPHOT and 450/850μm SCUBA observations (Alton et al. 1998a,b). Valentijn and van der Werf (1999b) have observed v=0 S(0) and S(1) $H_2$ emission throughout the disk of the edge-on galaxy NGC 891. Remarkably $H_2$ is detected out to ≥11 kpc galacto-centric radius, implying the presence of warm (150-230 K) $H_2$ clouds that may represent a significant fraction of the total molecular content of the galaxy. Based on the line shapes of the two $H_2$ lines at the outermost positions, Valentijn and van der Werf argue that there must be a second, cooler $H_2$ component (~80 K) that would then have to contain 4 to 10 times the mass estimated from CO and HI observations. If confirmed, such a molecular component (although much warmer than originally proposed by Pfenniger et al) could make a significant contribution to the dynamical mass at that radius.

## 2.3 Early Type Galaxies

IRAS observations have shown that a significant fraction of early type galaxies has detectable infrared emission (e.g. Knapp et al 1989, 1992). With ISO progress has been made in better understanding the origin of the emission. One key question is whether the mid-IR emission originates mainly in the photospheres and shells of cool asymptotic giant branch (AGB) stars, or in the interstellar medium, or whether an active galactic nucleus plays a role. Several studies with ISOCAM and ISOPHOT in more than 40 early type galaxies have addressed this issue (Madden et al 1999, Vigroux et al 1999, Knapp et al 1996, Fich et al 1999, Malhotra et al 1999b). Exploiting the imaging capability of ISOCAM at different wavelengths these studies show that all three components play a role. Madden et al (1999) find that the 4.5/6.7μm and 6.7/15μm colors of about 60% of their sample galaxies are consistent with a largely stellar origin for the mid-IR emission. The other galaxies are dominated by UIB emission (17%) or an AGN (22%). Half of the sample exhibits a mid-IR excess due to hot interstellar dust at some level. The interstellar dust emission is associated with the optical dust lane in NGC 5266 (Madden et al 1999) and in Cen A (Mirabel et al 1999).

The detection of UIB features in early type galaxies is of general significance, since the radiation field from the old stellar population is too soft to excite the UIB emission with UV photons (2.1.1). Either lower energy photons in the visible band are sufficient to excite the mid-IR features (Boulade et al 1996, Uchida et al 1998, Vigroux et al 1999, Madden et al 1999), or there has to be a young stellar component producing a stronger UV radiation field. Evidence for the latter explanation comes from the detection of 158μm [CII] (and 63μm [OI]) far-IR line emission in NGC 1155, NGC 1052 and NGC 6958 (Malhotra et al 1999b), and from Cen A (Madden et al 1995, Unger et al 1999). Excitation of [CII] line emission (3.1, 3.2.3) definitely requires ≥11.3 eV photons and the old stellar population in these galaxies cannot account for the required far-UV radiation field (Malhotra et al 1999b). A fair amount of the [CII] emission may come from a diffuse neutral HI medium. However, an additional obscured source of UV photons is required and is probably associated with massive star formation in dense interstellar clouds. The [OI]/[CII] ratio in NGC 1155 indicates that the radiation field is about one hundred times greater than that in the solar neighborhood and the gas density is ~100 cm$^{-3}$, characteristic of moderately dense PDRs associated with molecular clouds (3.2.3). Most of the [CII] emission in Cen A originates in the central dust disk where ISOCAM observations provide strong evidence for ongoing star forming activity (Mirabel et al 1999).

# 3. ACTIVE GALAXIES

## 3.1 Analytical Spectroscopy and Spectro-photometry

The 2 to 200μm band accessible to the ISO spectrometers (6.5 octaves!) contains a plethora of atomic, ionic and molecular spectral lines spanning a wide range of excitation potential, along with various solid-state features from dust grains of different sizes (2.1.1, Figure 3). These lines sample widely different excitation/ionization states and are characteristic tracers of different physical regions: photodissociation regions (PDRs[4]), shocks, X-ray excited gas, HII regions photoionized by OB stars and 'coronal' gas photoionized by a hard central AGN source (see Genzel 1992 for a review).

---

[4] PDRs are the origin of much of the infrared radiation from the interstellar medium (ISM). PDRs are created when far-UV radiation impinges on (dense) neutral interstellar (or circumstellar) clouds and ionizes/photodisscoiates atoms and molecules. The incident UV (star) light is absorbed by dust grains and large carbon molecules (such as PAHs) and is converted into infrared continuum and UIB features. As much as 0.1-1% of the absorbed starlight is converted to gas heating via photoelectric ejection of electrons from grains or UIBs (Hollenbach & Tielens 1997, Kaufman et al 1999).

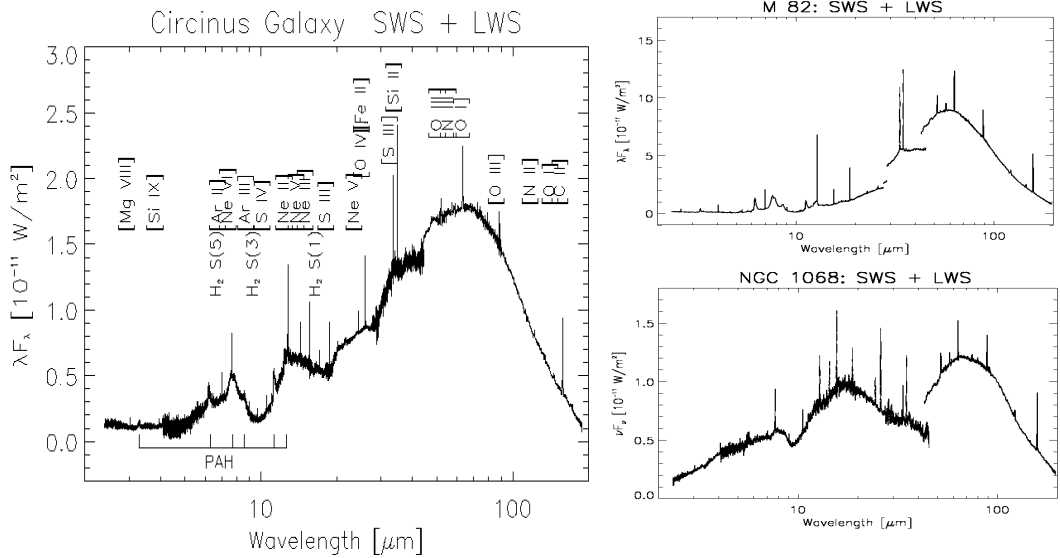

**Fig.3**

As an example of the rich spectra that were obtained with ISO we show in Fig.3 the combined SWS/LWS spectrum of the Circinus galaxy which displays the entire set of spectral characteristics just discussed. Circinus is the closest Seyfert 2 galaxy (D=4 Mpc), and has an active circum-nuclear starburst (Moorwood et al 1996, Moorwood 1999, Sturm et al 1999b).

*3.1.1 Molecular Spectroscopy*

ISO has opened the realm of external galaxies for molecular infrared spectroscopy. Pure $H_2$ rotational emission lines have been detected in a wide range of galaxies, from normal to ultra-luminous (e.g. Rigopoulou et al. 1996a, 2000, Kunze et al 1996, 1999, Lutz et al 1999b, Moorwood et al 1996, Sturm et al 1996, Valentijn et al 1996, Valentijn & van der Werf 1999a). The lowest rotational lines (28μm S(0) and 17μm

S(1)) originate in warm (90-200 K, Valentijn & van der Werf 1999a,b) gas that constitutes up to 20% of the total mass of the (cold) molecular ISM in these galaxies (Kunze et al 1999, Rigopoulou et al 2000). High rotation lines (S(5), S(7)) and ro-vibrational lines come from a small amount of much hotter gas ($\geq$1000 K). It is likely that the $H_2$ emission arises from a combination of shocks, X-ray illuminated clouds (in AGNs) and photodissociation regions. OH, CH and $H_2O$ lines are seen in emission or absorption (or a combination) in a number of gas rich starburst, Seyfert and ultra-luminous galaxies (Colbert et al 1999, Bradford et al 1999, Fischer et al 1997, 1999, Spinoglio et al 1999, Skinner et al 1997, Kegel et al 1999). The observed OH emission can generally be accounted for by infrared pumping through rotational and ro-vibrational transitions in sources with a warm infrared background source. The ISO SWS/LWS spectroscopy solves the long standing puzzle of the pumping mechanism for intense radio 'mega-maser' OH emission in luminous infrared galaxies (Baan 1993). In the megamaser galaxies Arp 220, IRAS 20100-4156 and 3Zw35 absorption is seen in cross-ladder rotational transitions (Fig.4), strongly supporting rotational pumping models for the masers (Skinner et al 1997, Kegel et al 1999). The location of the OH masers, therefore, must be in front of the far-IR continuum source on kpc-scales. The OH megamasers originate in the extended starburst region, and not in the circum-nuclear environment (Skinner et al 1997). The far-IR molecular line spectra of the ultra-luminous galaxies Arp 220 (Fig.4) and Mrk 231 are remarkably similar to those of the Galactic center molecular cloud complex/star forming region SgrB2 (Fig.4, Cox et al 1999, Fischer et al 1997, 1999), indicating that the properties of the entire molecular ISM in Arp 220 (scale 1 kpc) are comparable to those in this dense Galactic cloud ($N(H_2) \geq 10^{24}$ cm$^{-2}$, $n(H_2) \sim 10^4$ cm$^{-3}$).

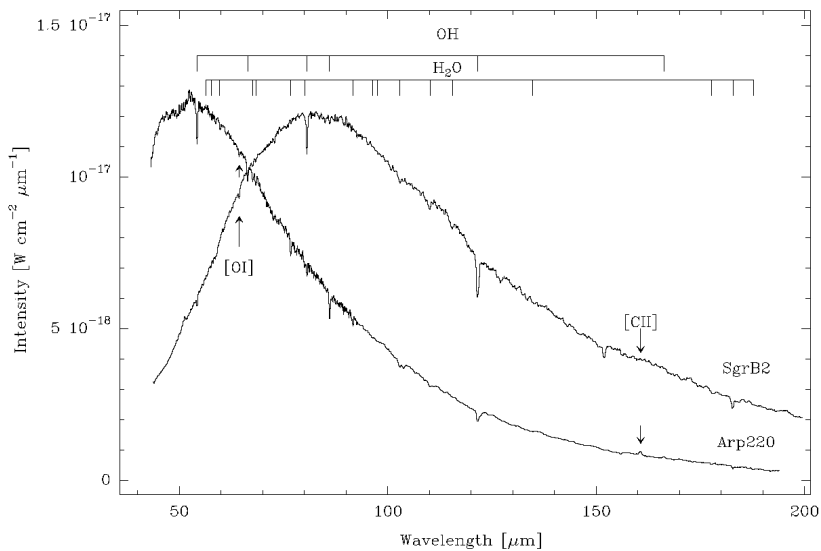

**Fig.4**

*3.1.2 Distinguishing AGNs and starbursts: 'ISO diagnostic diagrams'*

One of the most powerful applications of the ISO (mid-) IR spectroscopy is as a tool for distinguishing between AGN and star formation dominated sources. The right insets of Fig.3 (from Sturm et al 1999b) show two examples; the starburst galaxy M82 and the nuclear region of the Seyfert 2 galaxy NGC 1068. The difference between the mid-IR spectra is striking, in agreement with earlier ground-based results (Roche et al 1991). M82 is characterized by strong, low excitation fine structure lines, prominent UIB features and a weak $\lambda \leq 10\mu$m continuum (Sturm et al 1999b). High excitation lines are absent or weak. In contrast NGC 1068 displays a fainter, but highly excited emission line spectrum ([OIV]/[NeII] and [NeV]/[NeII]$\geq$1, Lutz et al 1999c), and no or weak UIB features, plus a strong mid-IR continuum (Genzel et al 1998, Sturm et al 1999b). The far-IR spectra of NGC 1068 and M82 (and Circinus) are more similar since star formation in the disks dominates the $\lambda\geq40\mu m$ SED in all three galaxies (Colbert et al 1999, Spinoglio et al 1999). The deep $10\mu m$ dip in the mid-IR SED of M82 (and other galaxies) has often been interpreted as due to absorption by silicate dust grains ($\tau_{9.7\mu m}\sim 2$ or A(V)~30). Sturm et al (1999b) have shown that the M82 spectrum can be fitted extremely well be the superposition of the (absorption free) spectrum of the reflection nebula NGC 7023 (exhibiting strong UIB features) plus a VSG continuum at $\lambda>10\mu m$. No silicate absorption is required. Instead the deep $10\mu m$ dip is caused by the strong UIB emission on either side of the silicate feature.

'Diagnostic diagrams' can empirically characterize the excitation state of a source (Osterbrock 1989, Spinoglio & Malkan 1992, Voit 1992). In the form employed by Genzel et al (1998) a combination of the $25.9\mu m$ [OIV] to $12.8\mu m$ [NeII] line flux ratio (or [OIV]/[SIII], or [NeV]/[NeII]) on one axis and of the UIB strength[5] on the other axis is used. Fig.5a shows that this 'ISO diagnostic diagram' clearly separates known star forming galaxies from AGNs. The three AGN templates that are located fairly close to the starbursts in Fig.5a (CenA, NGC 7582 and Circinus, Mirabel et al 1999, Radovich et al 1999, Moorwood et al 1996) are well known to contain circum-nuclear starbursts; correction for the star formation activity would move these sources to the upper left (marked by arrows). The diagram is insensitive to dust extinction if numerator and denominator in both ratios are affected by a similar amount of extinction. Lutz et al (1998b) find low level [OIV] emission ([OIV]/[NeII]~$10^{-2}$) in a number of starburst galaxies. The [OIV] emission in M82 is spatially extended and tracks galactic rotation. It cannot come from low level AGN activity. Fast, ionizing shocks in galactic 'super' winds are the most likely sources of this low level, high excitation gas (Lutz et al 1998b, Viegas et al 1999) but WR stars may play a role in some objects (Schaerer & Stasinska 1999).

---

[5] The $7.7\mu m$ (UIB) 'line to continuum' ratio or 'strength' is the ratio of the peak flux density in the UIB feature (continuum subtracted) to the underlying $7.7\mu m$ continuum flux density. The latter is computed from a linear interpolation of the continuum between $5.9\mu m$ and $11.2\mu m$. While the $11.2\mu m$ continuum is often affected by silicate absorption and various emission features longward of $11\mu m$, $7.7\mu m$ is close enough to the fairly clean $5.9\mu m$ region that a simple linear extrapolation is justified as a simple and fairly robust estimate of the $7.7\mu m$ continuum.

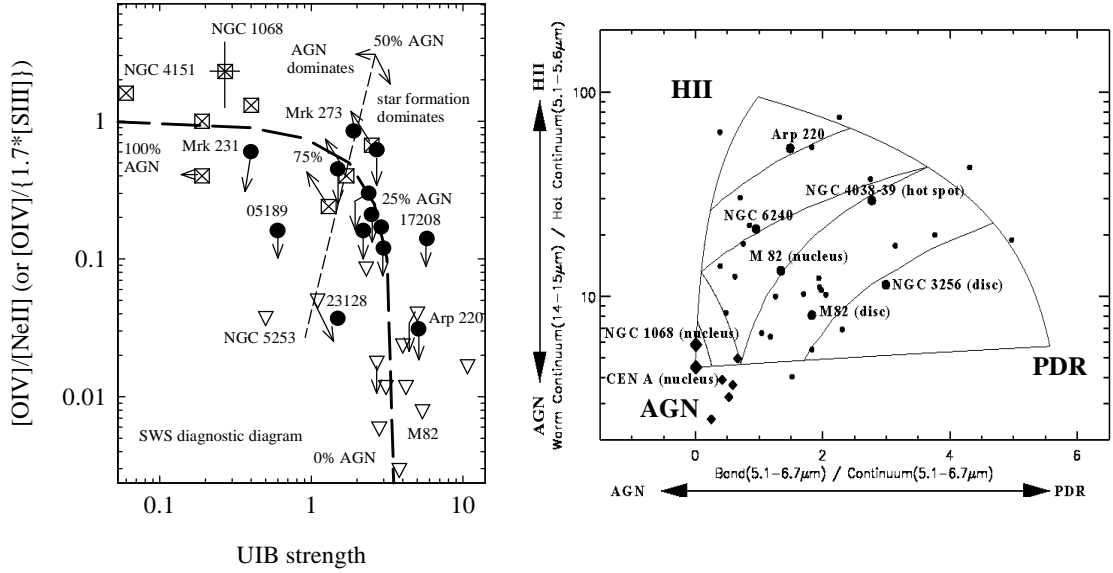

**Fig.5**

Based on ISOCAM CVF spectra of the central part of M17 (an HII region), of the nuclear position of Cen A (an AGN, Mirabel et al 1999) and the reflection nebula

NGC 7023 (a PDR, D Cesarsky et al 1996 a) Laurent et al (1999) propose another diagnostic diagram. It is based on the 15µm/6µm continuum ratio on one axis and the UIB/6µm continuum ratio on the other axis. Laurent et al demonstrate that this mid-IR diagnostic diagram also distinguishes well between AGNs and starbursts (Fig. 5b).

One example of the value of the techniques proposed by Laurent et al (1999) is the study of Centaurus A, the closest radio galaxy. In the visible band Cen A is a giant elliptical galaxy with a prominent central dust lane which has been interpreted already 45 years ago to be the result of a merger between the elliptical and a small gas rich galaxy (Baade & Minkowski 1954). The central AGN is hidden in the optical band and its presence can only be directly seen from the prominent double lobed, radio and X-ray jet system (Schreier et al. 1998). The ISOCAM mid-IR observations penetrate the dust and reveal a strong and compact, hot dust source associated with the central AGN, as well as a bi-symmetric structure of circum-nuclear dust extending in the plane of the dust lane (Fig.6a, Mirabel et al 1999). The mid-IR spectrum toward the nuclear position is characteristic of an AGN as described in the last sections (Genzel et al 1998, Mirabel et al 1999, Laurent et al 1999, D Alexander et al 1999). In contrast the off-nuclear spectra are characteristic of star forming regions (Madden et al 1999a, Unger et al 1999). Mirabel et al interpret the bi-symmetric dust structure as a highly inclined ($i=72^o$) barred spiral. Its cold dust content is comparable to a small spiral galaxy. The gas bar may funnel material from kpc-scales toward the circum-nuclear environment ($10^2$ pc scale), from where it may be transported further by a circum-nuclear bar/disk inferred from near-IR polarization observations. As such Cen A is a beautiful nearby example of how a merger can convert an (old) spheroid into a bulge+disk system.

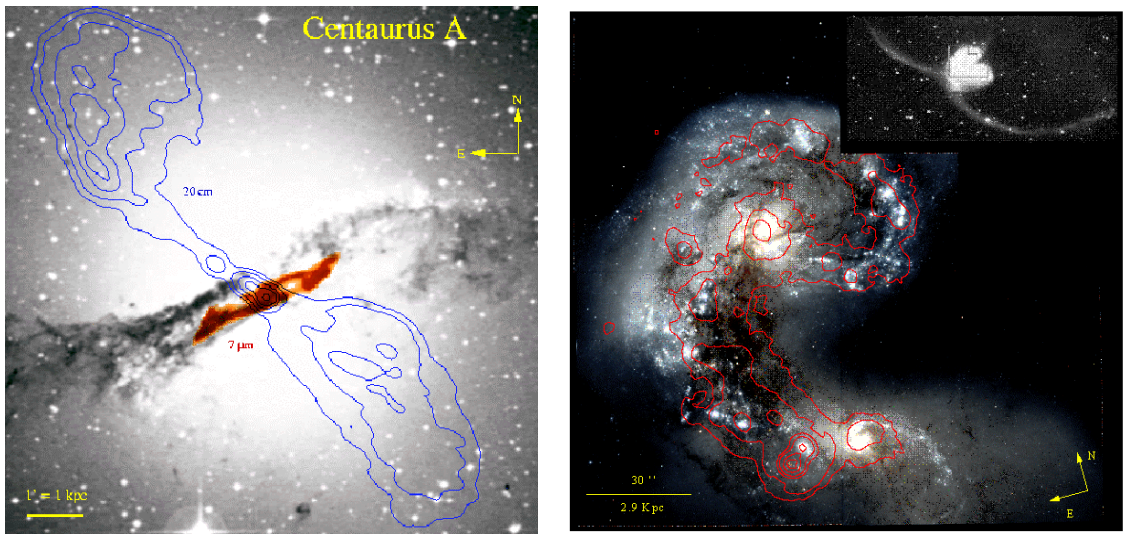

**Fig.6**

## 3.2 Starburst Galaxies

### *3.2.1 The dark side of star formation*

The beautiful HST image of the famous 'Antennae' system of colliding galaxies
(NGC 4038/39: Fig.6b, from Whitmore et al 1999) summarizes one of the paradigms
about active star formation on galactic scales: spectacular fireworks of star formation
with hundreds of young star clusters are triggered by strong gravitational
disturbances. The interaction/collision and eventual merging of two gas rich galaxies
is the most effective such perturbation. Bar driven inflow is another. ISOCAM studies
of several samples of interacting, barred and collisional ring galaxies confirm this paradigm (Appleton et al 1999, Charmandaris et al 1999, Roussel et al 1999b, Wozniak et al 1999). The optical/UV data in the Antennae tell a fascinating story with evidence for a large age spread in star cluster populations. There have been several phases of active star formation since the first interaction ~500 million years ago. Yet the UV/optical data do not reveal the whole story. Superposed on the HST image in Fig.6b are the contours of 15µm mid-IR emission (Mirabel et al 1998). The 15µm map is an (approximate) tracer of the entire infrared emission that contains ~80% of the bolometric luminosity of the system. About half of the entire luminosity emerges from the optically 'dark' interaction region between the two galaxy nuclei. This region also contains most of the molecular gas (Stanford et al 1990, Lo et al 1999). A compact source at the southern edge of the interaction region accounts for ~15% of the entire luminosity. Yet on the HST image there is only a faint and very red, compact cluster at this position. This knot is the site of the most recent star formation activity in the Antennae (see section 3.2.3, Vigroux et al 1996, Kunze et al 1996).

Another example is the luminous merger Arp 299 (NGC 3690/IC694, Gehrz et al 1983, Satyapal et al 1999). In this source, as in the Antennae, more than 50% of the star forming activity orginates outside the (bright) nuclei, much of it in a highly obscured, compact knot 500pc south-east of the nucleus of NGC 3690 (Gallais et al 1999). Finally, Thuan et al (1999) have observed with ISOCAM the blue compact dwarf galaxy SBS0335-052 which has the second lowest metallicity kown in the local Universe ($Z=Z_\odot/41$). They find that this starburst system is remarkably bright in the mid-IR ($L_{12\mu m}/L_B=2$). The mid-IR spectrum can be fitted by a ~260K blackbody with strong 9.7/18µm silicate absorption features and no UIB emission. A significant fraction of the starburst activity of SBS0335-052 thus must be dust enshrouded, potentially a very important clue for the understanding of low metallicity starbursts in the Early Universe (4).

*3.2.2 Properties and evolution of starbursts*

Dusty starburst galaxies constitute an important class of objects (c.f. Moorwood 1996). About 25% of the high-mass star formation within 10 Mpc occurs in just 4 starburst galaxies (Heckman 1998). Key open questions of present research relate to the form of the initial stellar mass function (IMF) in starbursts and to the burst evolution. ISO has contributed to these issues mainly by studying the content of the most massive stars in dusty starbursts through nebular spectroscopy. Following pilot studies by Fischer et al (1996), Kunze et al (1996), Lutz et al (1996a) and Rigopoulou et al (1996a), Thornley et al (1999) have carried out an SWS survey of [NeIII]/[NeII] line emission in 27 starburst galaxies, with a range of luminosities from $<10^8$ to $>10^{12}$ $L_\odot$. The excitation potentials of $Ne^+$ and $Ne^{++}$ are 22 and 41 eV and the two lines are very close in wavelength and have similar critical densities. The Neon line ratio is a sensitive tracer of the excitation of HII regions and of the OB stars photoionizing them. The basic result for the starburst sample is that HII regions in dusty starburst galaxies on average have low excitation (left inset of Fig.7). The average [NeIII]/[NeII] ratio in starburst galaxies is a factor of 2 to 3 lower than in individual Galactic compact HII regions. The SWS work establishes as a fact and puts on a safe statistical footing what was already suggested by earlier ground-based near-IR results (e.g. Rieke et al 1980, Doyon et al 1994, Doherty et al 1995).

Thornley et al (1999) have modelled the HII regions as ionization bounded gas clouds photoionized by central evolving star clusters. The cluster SED (as a function of IMF and evolutionary state) is used as the input for photoionization modelling of the nebular emission. The computed Ne-ratios (for a Salpeter IMF with different upper mass cutoffs) as a function of burst age $t_b$ and burst duration $\Delta t_b$ are plotted in Fig. 7 (lower right). The basic result is that on average stars more massive than about 35 to 40 M are not present in starburst galaxies, either because they were never formed (upper mass cutoff) or because they already have disappeared because of aging effects.

An upper mass cutoff in the intrinsic IMF is difficult to reconcile with a growing body of direct evidence for the presence of very massive stars (≥100 M ) in nearby starburst templates (Galactic center: Krabbe et al 1995, Najarro et al 1997, Serabyn et al 1998, Figer et al 1998, Ott et al. 1999, R136 in 30 Doradus: Hunter et al 1995, Massey & Hunter 1998, NGC 3603: Drissen et al 1995, Eisenhauer et al 1998). In more distant HII region galaxies the 4696 Å HeII emission line feature signals the presence of Wolf-Rayet stars of mass ≥60 M (e.g. Conti & Vacca 1994, Gonzalez-Delgado et al 1997). The low average excitation in starbursts is thus more plausibly caused by aging (Rieke et al 1993, Genzel et al 1994). This conclusion is also supported by the large spread and substantial overlap of the Galactic and extragalactic Ne-ratios (Fig.7). The overlap region of the Antennae and the near-nuclear knot in NGC 3690 have [NeIII]/[NeII]~1, comparable to nearby HII regions. Precisely in these two regions most of the flux (3.2.1) comes from a single compact (and probably young) region. Everywhere else in the Antennae the Ne-ratio is much lower (Vigroux et al 1996) and at the same time the stellar cluster data (Whitmore et al 1999) indicate greater ages. Similar results emerge if, instead of the Ne-ratio, far-infrared lines ratios (Colbert et al 1999), or the ratio $L_{bol}/L_{Lyc}$ are used as constraints (Thornley et al 1999).

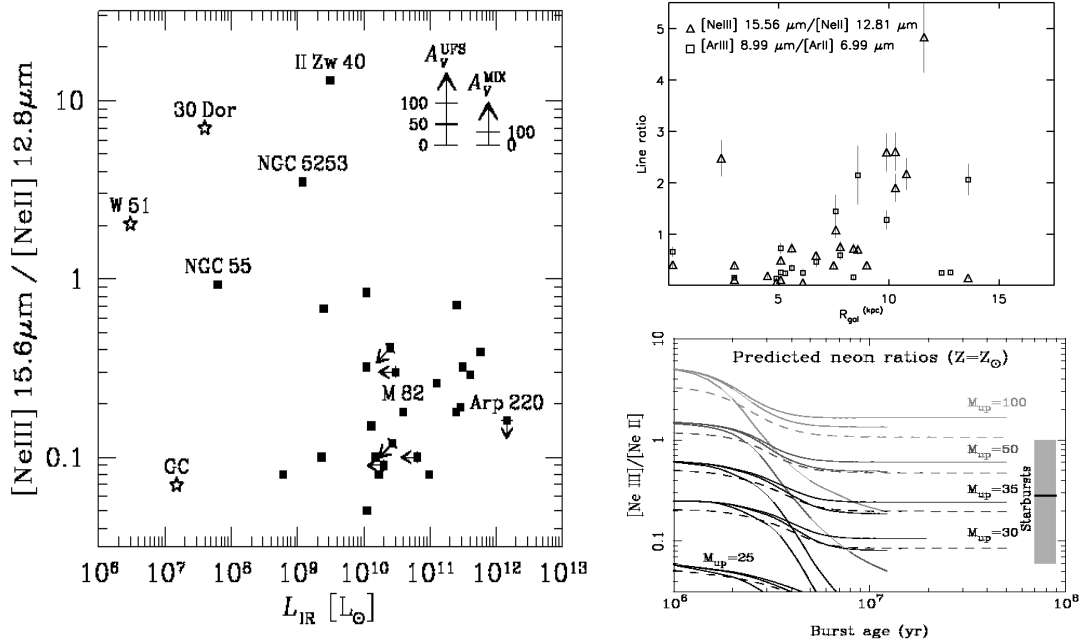

**Fig.7**

If stars of masses 50-100 M are initially formed in most galaxies, the inferred burst durations must be less than 10 Myr (Fig.7). Such short burst time scales are surprising especially for luminous, distant systems (3.4.4). The 'communication' time is a few million years or more. It appears that supernovae and stellar winds disrupt the ISM and prevent further star formation as soon as the first generation of O stars have formed and evolved. Starbursts must induce large negative feedback. Starburst models (Thornley et al 1999, Leitherer and Heckman 1995) as well as observations of galactic superwinds (Heckman et al 1990) indicate that 1 to 2% of the bolometric luminosity of starbursts emerges as mechanical energy in superwinds. If this mechnical energy input removes interstellar gas from the potential well with ~10% efficiency, the starburst is choked after the 'dispersal' time $t_d$. For typical gas masses and starburst parameters $t_d$ is $\leq 10^7$ years. The burst durations thus appear to be significantly smaller than the gas consumption time scales, $t_{gc}$~ 5 to $10 \times 10^7$ yrs. Near-IR imaging spectroscopy in M82, IC 342 and NGC 253 indicates that in the evolution of a given starburst system, there are several episodes of short burst activity (Rieke et al 1993, Satyapal et al 1997, Böker et al 1997, Engelbracht et al 1998, Förster-Schreiber et al 1999).

While the scenario just given is qualitatively quite plausible, the small values of the burst time scales indicated by the modelling of Thornley et al (1999) seem unphysical. In reality burst time scales are probably somewhat larger (~$10^7$ years). The Galactic center starburst region has been analyzed through a direct stellar census (Krabbe et al 1995, Najarro et al 1997). The Ne-ratio computed from the starburst model ($t_b$~$7 \times 10^6$, $\Delta t_b$~$4 \times 10^6$ years, log U=-1) is more than an order of magnitude greater than the observed one (Lutz et al 1996b). The models predict that the Galactic center, ionizing UV continuum should be dominated by O stars and hot Wolf-Rayet stars. The data show, however, that more than half of the ionizing flux comes from a population of relatively cool (20,000–30,000 K) blue supergiants (Najarro et al 1997). It would thus appear that the tracks and stellar properties adopted for the Galactic center are not appropriate (Lutz 1999). Likely culprits could be the effects of metallicity and of the dense stellar winds on the UV SEDs of the massive stars. The Ne-ratio depends strongly on metallicity, as is directly demonstrated by the much higher excitation of II

Zw 40 and NGC 5253 (Z=0.2-0.25 Z ) and of 17 HII regions in the SMC/LMC observed by Vermeij & van der Hulst (1999). The average Ne-abundance in the starburst sample is ~1.7 times solar. Yet the models of Thornley et al (1999) are for solar metallicity. Furthermore there appears to be a Galactocentric gradient in the Ne-ratios of Cox et al (1999), again signalling dependence on metallicity. A similar discrepancy between (high excitation) current stellar models and (low excitation) observed nebular emission lines has also been found for Wolf-Rayet stars (Crowther et al 1999a), suggesting that their UV-spectra are softer than considered so far (Hillier & Miller 1998). Another area of concern is the impact on the interpretation of spatial variations in the various tracers that cannot be properly taken into account by single aperture spectroscopy. Crowther et al (1999b) report SWS and ground-based infrared spectroscopy of the low-metallicity starburst galaxy NGC 5253. While [SIV] and Brα come from a very compact (3") region centered on the nucleus, the [NeII] emission is much more extended. There is a high excitation ($T_{eff}$>38,000 K) young (t=2.5 Myr) nuclear burst, surrounded by a lower excitation ($T_{eff}$~35,000K) older (5 Myr) one. Thornley et al have raised the possibility that some of the [NeII] flux in their starburst galaxies may come from a more diffuse, low ionization parameter zone, thus lowering the effective Ne-ratio in the SWS aperture.

### 3.2.3 The [CII] line as global tracer of star formation activity in galaxies

Observations with the Kuiper Airborne Observatory (KAO), the COBE satellite and balloons have established that the 158 μm [CII] emission line commonly is the strongest spectral line in the far-IR spectra of galaxies (Crawford et al 1985, Stacey et al 1991, Lord et al 1996, Wright et al. 1991, Mochizuki et al 1994). The [CII] line traces PDRs as well as diffuse HI and HII regions. It should be an excellent tracer of the global galactic star formation activity, including that of somewhat lower mass (A+B) stars (Stacey et al 1991).

There have been a number of LWS studies of the [CII] (and [OI]) lines. Malhotra et al (1997, 1999a), Fischer et al (1996, 1999), Lord et al (1996,2000), Colbert et al (1999), Helou et al (1999) and Braine and Hughes (1999) have observed 70 star forming galaxies over a wide range of activity level. Smith and Madden (1997) and Pierini et al (1999) have studied 21 Virgo cluster galaxies of normal and low star formation activity. Luhman et al (1998, and priv.comm.) and Harvey et al (1999) have observed 14 ultra-luminous galaxies. Figs.8a and b summarize these results (along with earlier work) and puts them in the context of PDR models. For most normal and moderately actively star forming galaxies, the [CII] line is proportional to total far-IR flux and contains between 0.1 and 1% of the luminosity (Fig.8 b, Malhotra et al 1997). In a plot[6] of the ratio of [CII] line flux to integrated far-IR continuum flux ($Y_{[CII]}$) as a function of the ratio of CO 1-0 line flux to FIR flux ($Y_{CO}$) most extragalactic (and Galactic) data are in good overall agreeement with PDR models for hydrogen densities of $10^{3..5}$ and radiation fields of $10^{2..4}$ times the solar neighborhood field.

However, for luminous Galactic HII regions and the most active star forming galaxies, including ULIRGs, the [CII]/FIR ratio plumets to <0.1%. Malhotra et al (1997, 1999a) find that there is an inverse correlation between $Y_{[CII]}$ and dust temperature and also between $Y_{[CII]}$ and star formation activity (as measured by the ratio of IR to B- band luminosities, Fig. 8b). The [CII] line is faint for the most IR-luminous galaxies (Luhman et al 1998, Fischer et al 1999). In the context of standard PDR models, one major reason for the drop of $Y_{[CII]}$ probably is the lower gas heating efficiency at high radiation fields (Malhotra et al 1997). For the high UV energy densities characteristic of ULIRGs and other active galaxies, the photoelectric heating efficiency is low. This is because the dust grains are highly positively charged and the UIB molecules (important contributors of the photoelectrons: Bakes and Tielens 1994) are destroyed. As a consequence the line emissivity of the PDR tracers drops sharply. Another factor may be that in such active galaxies the pressure and gas density of the ISM is significantly greater than in normal galaxies and the $^2P_{3/2}$ level of the ground state of $C^+$ is collisionally deexcited. The location of the ULIRGs in Fig.8a can be qualitatively understood if gas densities exceed $10^5$ cm$^{-3}$. An increase of the 63μm [OI]/[CII] line ratio with density is, in fact, observed in the US Key Project sample (Fig.8 b, Malhotra et al 1999a) but can only account for part of the decrease in $Y_{[CII]}$.

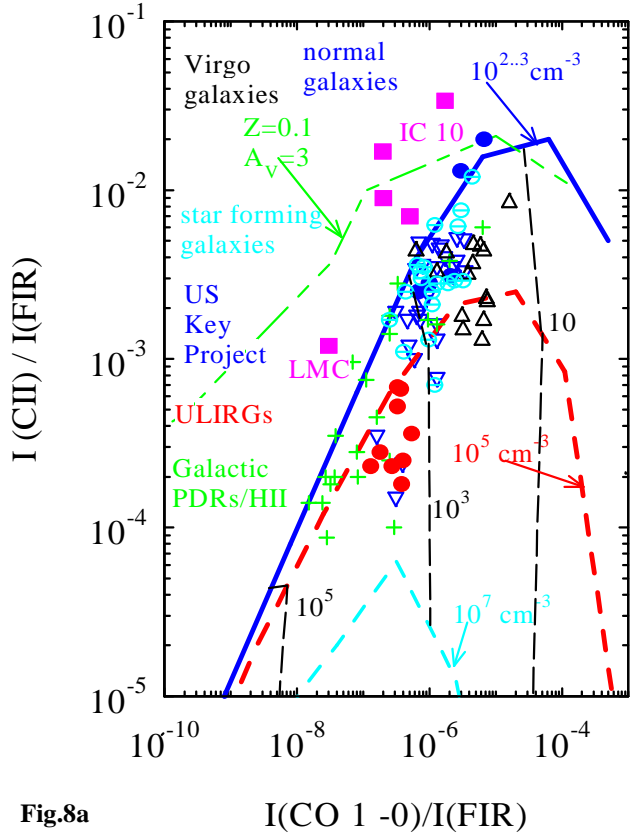

**Fig.8a**    I(CO 1 -0)/I(FIR)

---

[6] This $Y_{[CII]}$-$Y_{CO}$ plot was first introduced by Wolfire et al (1990) to remove the dependence on the PDR filling factor in the data and allows a direct comparison with models. These models assume that all three tracers come from the same regions, an assumption that is not always fulfilled.

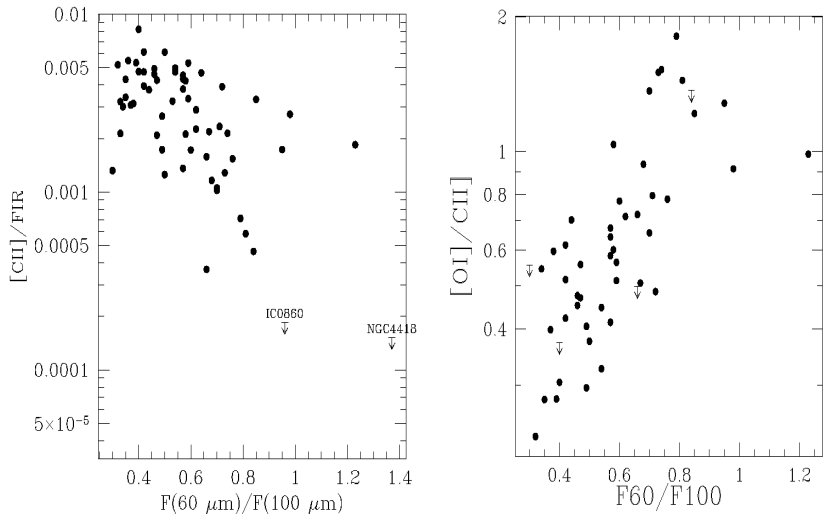

**Fig.8b**

Another important effect is self-absorption in the lines. In the yet more extreme ULIRGs of the Luhman et al sample both the 63μm [OI] and the 145μm [OI] lines are weak (or even in absorption) in Arp 220 (Fig.4) and Mrk 231 (Fischer et al 1997, 1999, Harvey et al 1999). The Sgr B2 star forming region in the Galactic center perhaps comes closest to the conditions in ULIRGs in terms of UV and FIR energy density, high gas density and large $H_2$ column density. In Sgr B2 both the [CII] and [OI] lines show (partial) self-absorption due to intervening cold or subthermally excited gas (Fig.4, Cox et al 1999). The self-absorption model, however, does not explain the lack of 145μm [OI] emission in Arp 220 (Luhman et al 1998) as its lower level is 228 K above the ground state and should not be populated in cold or diffuse gas. Fischer et al (1999) propose that the radiation field in ULIRGs is soft, perhaps as a result of an aging burst. Nakagawa et al (1996) give the same explanation for the relative weakness of the [CII] emission in the extended Galactic center region. The moderately strong [NeII] and [SIII] lines (requiring substantial Lyman continuum luminosity) detected in the same sources, however, makes this explanation fairly implausible.

Low metallicity galaxies often exhibit unusually intense [CII] emission, as compared to the CO mm-emission (Stacey et al 1991, Mochizuki et al 1994, Poglitsch et al 1995, Madden et al 1997, but see Israel et al. 1995, Mochizuki et al 1998). They lie in the upper left in the $Y_{[CII]}$-$Y_{CO}$ plot (Fig.8a). Low metallicity by itself cannot account for this effect but the combination of low metallicity with small, cloud column densities can (Poglitsch et al 1995, Pak et al 1998). The thin dashed curve in the upper left of Fig 8 is model for an $A_V$=3, Z=0.1 Z cloud (Kaufman et al 1999).

In the other extreme of low star forming activity encountered in the Virgo sample, the [CII]/FIR ratio also is lower than in moderately active galaxies (Smith and Madden 1997, Pierini et al 1999). Again this is expected in PDR models (Fig.8a) as a result of inefficient heating of the PDRs at low energy density, of a lower ratio of UV to FIR energy density and of lower gas heating efficiency in a lower density, 'HI'-medium (Madden et al 1993, 1997).

In summary then, the ISO LWS results generally confirm the predictions of PDR theory (Tielens & Hollenbach 1985, Hollenbach & Tielens 1997, Kaufman et al 1999) and support the interpretation that the [CII] line is a good tool to trace global star formation in normal and moderately active galaxies. The decrease of the [CII]/FIR ratio in very active galaxies and ULIRGs came as a surprise. In retrospect some of that should have been expected on the basis of the theoretical predictions. Much hope has been placed on future applications of the [CII] line as a tracer of global star formation at high redshifts. The ISO data cast some doubt that these expectations will be realized for the most luminous objects unless low metallicity helps out.

## 3.3 Seyfert Galaxies and QSOs

### *3.3.1 UIB features in Seyfert galaxies: confirmation of unified models*

Clavel et al (1998) have used the UIB features as a new tool for testing unified schemes[7] in Seyfert 1 and 2 galaxies. The basic idea is that hot dust at the inner edge of the postulated circum-nuclear dust/gas torus surrounding the central AGN (Antonucci 1993) reemits the absorbed energy from the central AGN as a thermal continuum in the near- and mid-IR. In the torus models of Pier and Krolik (1992, 1993) and Granato and Danese (1994) the torus is optically thick even at mid-infrared wavelengths. The hot dust emission is predicted to be much stronger in Seyfert 1 than in Seyfert 2 galaxies, since the mid-IR continuum is observing angle dependent due to the torus blockage/shadow. The facts that the infrared energy distributions of AGNs (Seyferts and QSOs) are fairly broad and the silicate absorption/emission is relatively weak in most of the objects in the Clavel et al sample (Fig.9) argues against very compact and very thick tori. The data are in better agreement with moderately thick and extended tori, or clumpy disk configurations, as already pointed out before the ISO mission by Granato et al (1997) and Efstathiou and Rowan-Robinson (1995).

---

[7] In unified models (e.g. Antonucci 1993) different types of AGNs are postulated to all harbor a central, accreting massive black hole surrounded by a broad line region. The AGN is proposed to be surrounded by an optically and geometrically thick gas/dust torus that anisotropically absorbs/shadows emission from the nuclear region. Broad-line Seyfert 1 galaxies (or QSOs) in this scheme are AGNs where the line of sight to the nucleus is not blocked. In narrow-line, Seyfert 2 galaxies (or radio galaxies) the line of sight to the nucleus is blocked by the intervening torus.

Clavel et al have carried out ISOPHOT-S spectrophotometry of a sample of 26 Seyfert 1-1.5 (<z>=0.036) and 28 Seyfert 1.8-2 (<z>=0.017) galaxies drawn from the CFA sample (Huchra & Burg 1992). Their main finding is that the equivalent width of the 7.7μm UIB feature is much stronger in Seyfert 2 galaxies than in Seyfert 1 galaxies (Fig. 9). The median 7.7μm UIB strength[5] of Seyfert 2 galaxies is ~4 times larger than for Seyfert 1s (7.7μm L/C=1.7±0.37 vs. 0.4±0.13). This difference is not due to Seyfert 1s having weaker UIB emission but due to their having a much stronger (red) continuum emission than Seyfert 2s (Fig.9). The distributions of UIB luminosities (and the ratios of UIB to far-IR luminosities) of the two Seyfert classes are similar (Fig.10). The UIB/far-IR ratio in Seyferts is also very similar to that in starburst galaxies (Fig.10).

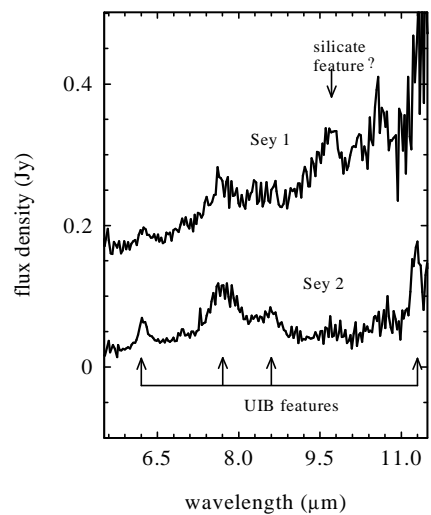

**Fig.9**

The findings of Clavel et al (1998) are in agreement with earlier ground-based, mid-IR photometry of several Seyfert samples (Maiolino et al 1995, Giuricin et al 1995, Heckman 1995). The results are in excellent agreement with the unified torus model discussed above if the UIB emission is orientation indendent. Spatially resolved ISOCAM CVF imaging in three nearby active galaxies (NGC 1068, Circinus and Cen

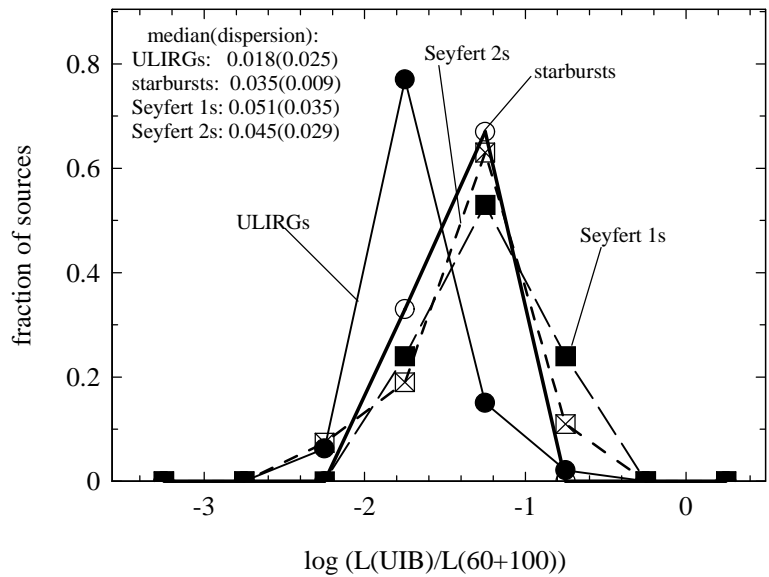

**Fig.10**

A) shows that the UIB emission comes from kpc-scale star forming regions in the disk, and not from the nucleus (Laurent et al 1999, Moorwood 1999). The similar UIB/far-IR ratios in Seyfert 1s, Seyfert 2s and starburst galaxies (Fig.10) suggest that the UIB luminosity (and the far-IR continuum) in Seyfert (and other) galaxies is proportional to the sum of star formation activity and emission from the diffuse ISM (3.3.2).

*3.3.2 The source of far-infrared luminosity in Seyfert galaxies and QSOs*

The IRAS data have shown that the infrared spectral energy distributions (SEDs) of Seyfert galaxies and (radio quiet) QSOs are dominated by thermal (dust) emission (Sanders et al 1989, Barvainis 1990, Pier and Krolik 1993, Granato and Danese 1994, Rowan-Robinson 1995, Granato et al 1997). ISOPHOT observations of the 2-200 µm SEDs of Seyferts, radio galaxies and QSOs have confirmed these conclusions and improved the quality and spectral detail of the data (Haas et al 1999, Wilkes et al 1999, Rodriguez Espinosa et al 1996, Perez Garcia et al 1998). As part of the European Central Quasar Program (~70 QSOs and radio galaxies) Haas et al (1998a, 1999) report ISOPHOT/IRAS SEDs, with additional 1.3mm points from the IRAM 30m telescope, for two dozen (mainly radio-quiet) PG QSOs. They conclude that all QSO in their sample, including the four radio loud ones, have substantial amounts of cool and moderately warm (20 to 60 K) dust radiating at $\lambda \geq 60$ µm, in addition to warmer (circum-nuclear) dust radiating in the mid-infrared. The derived dust masses ($10^{7\pm1}$ M☉) are typical for the total dust masses in gas rich, normal galaxies, suggesting that the $\geq 60$µm emission is probing the large scale disks of the host galaxies. Andreani (1999) has observed a complete subset of 34 QSOs from the Edinburgh and ESO QSO surveys that sample the bright end of the (local) QSO luminosity function. From her 11-160µm ISOPHOT observations she concludes that the mid-IR and far-IR fluxes are poorly correlated, as are the far-IR and blue-band fluxes. In contrast, the 60,100 and 160µm fluxes seem to be well correlated. This finding suggests that the far-IR emission in QSOs is a distinct physical component that may not be physically related to the shorter wavelength emission. Wilkes et al (1999) report the first results of the US Central QSO program that contains another 70 objects and extends to z=4.7. These data confirm as well the presence of thermal dust emission with a wide range of temperatures. The objects selected in that survey extend to the high luminosity tail of the high-z QSO population with infrared luminosities of several $10^{15}$ L . Van Bemmel et al (1999) report observations of matched pairs of QSOs and radio galaxies (radio power, distance etc.). They find that QSOs are actually more luminous far-IR sources than radio galaxies, contrary to simple unification schemes.

The key question that needs to be answered next is the nature of the energy source(s) powering the IR emission: direct radiation from the central AGN, or distributed star formation in the host galaxy. The near- and mid-IR emission ($\leq 30\mu m$) is very likely re-radiated emission from the AGN accretion disk (the 'big blue (or EUV) bump', 3.3.3). The $\lambda \leq 30\mu m$ SEDs can be well matched with the Pier and Krolik (1992, 1993) and Granato and Danese (1994) models of AGN heated dusty tori (scale size ~100 pc), with an additional component of somewhat cooler dust.

More difficult is the answer to the question of whether the $\lambda \geq 30\mu m$ emission is re-radiated AGN luminosity as well. A compact and thick torus, as proposed by Pier and Krolik (1992), for instance, definitely does not produce a broad enough SED to explain the far-IR emission. A clumpy, extended and lower column density torus (Granato et al 1997), a warped disk (Sanders et al 1989) or a tapered disk (Efstathiou & Rowan-Robinson 1995) are more successful in qualitatively accounting for the observed broad SEDs. However, even these models work only marginally for a quantitative modeling of the emission at $\lambda \geq 100\mu m$. For the CfA sample the correlation between $60+100\mu m$ band luminosity and (extinction corrected) [OIII] luminosity is not impressive for Seyfert 1s and outright poor for Seyfert 2s.

For QSOs no clear answer to the nature of the far-IR continuum has emerged yet. On the one hand, Sanders et al (1989) conclude that the far-IR emission in PG quasars is mainly due to AGN re-radiation. On the other hand, Rowan-Robinson (1995) and Haas et al (1999) argue that the far-IR emission is caused by star forming activity in the QSO hosts. In the Sanders et al model a warped disk could intercept at least 10% of the luminosity of the central AGN. In their sample of about ~50 radio quiet PG quasars with available IRAS luminosities the average ratio of total infrared to total UV+visible ('Big Blue (or EUV) Bump') luminosities is about 0.4. The $\lambda \geq 30\mu m$ luminosity is about half of the total infrared luminosity, thus requiring that about 15% of the nuclear luminosity be absorbed and re-radiated at $10^2$ to $10^3$ pc from the AGN. This may be possible if substantial warps are present on that scale. In contrast Rowan-Robinson (1995) cites the cases of three PG QSOs (0157+001, 1148+549, 1543+489) and the Seyfert 1/ULIRG Mrk 231 where the far-IR luminosity exceeds the optical+UV luminosity, a result that is not possible in the re-radiation scenario (see also 3.4.5). An interpretation of the far-infared emission in terms of star formation is also favored by the far-IR/radio relationship in Seyferts and radio quiet QSOs. Colina and Perez-Olea (1995, and references therein) show that the ratio of $60+100\mu m$ IRAS luminosity to 5 GHz radio luminosity in these objects is in excellent agreement with the ratio found in star forming spirals of a wide range of luminosities.

For Seyfert galaxies a clearer picture is emerging. For the CfA Seyfert sample of Clavel et al (1998) the median ratio of 60+100µm luminosity to B-band luminosity is 0.9 for Seyfert 1s and 1.6 for Seyfert 2s. Assuming that about 30% of the total UV+visible luminosity is contained in the B-band (as in the PG QSOs of Sanders et al 1989), the fraction of bolometric luminosity emerging in the far-IR in the Clavel et al Seyferts is about 40%. This is probably too large for a re-radiation scenario. Perez Garcia et al (1998) have investigated the infrared spectral energy distribution of 10 Seyfert galaxies with a combination of IRAS and ISOPHOT data. They decomposed the 4-200µm spectra into a sum of blackbodies with $\lambda^{-2}$ emissivities. In 9 of the 10 galaxies studied the infrared spectrum decomposes into three components each with a similar, narrow range of temperatures. A 110-150 K component dominates the mid-infrared. The far-IR emission comes from a combination of a 40-50 K (30 to 100µm) and a 10-20 K (150 to 200µm) component. Perez Garcia et al conclude from the similarity of the temperatures of the three components that they represent well separated spatial regions (compact torus, star forming regions and diffuse ISM), rather than a temperature range in a single physical component (i.e. a circum-nuclear torus or warped disk). In this interpretation the 40-50 K temperature component dominating the 60+100µm IRAS band is a measure of star formation in the disk of the Seyfert galaxies. Finally the finding (Fig.10) that the ratio of UIB luminosity to 60+100µm luminosity in the Clavel et al (1998) Seyferts is essentially the same as in starburst galaxies also strongly supports an interpretation of the far-IR emission in terms of (disk) star formation.

### *3.3.3 Reconstructing the Big Blue (EUV) Bump*

In the standard paradigm AGNs are powered by accretion onto massive black holes. From basic theoretical considerations about the size scale of the emitting region and the energy released one expects that accretion disks emit quasi-thermal radiation with a characteristic temperature of a few $10^5$ K (~40 eV). The intrinsic SEDs of AGNs cannot be directly observed between the Lyman edge and a few hundred eV, however, due to Galactic and intrinsic absorption. In Seyfert 1s and QSOs observations shortward (X-rays) and longward (UV) of the critical region indicate that there is a break in the continuum slope. The data are broadly consistent with the existence of an emission peak (the 'Big Blue Bump' (BBB)) in the extreme UV (EUV, Malkan and Sargent 1982, Sanders et al 1989, Walter et al 1994, Elvis et al 1994).

Observations of emission lines from highly excited species enable a different approach to the study of the EUV spectrum of AGNs. These 'coronal' lines (Oke & Sargent 1968) originate in the NLR and can thus be detected even in type 2 sources, yet they may (and should in standard unified schemes) probe the intrinsic ionizing continuum. Coronal lines are probably excited by photoionization (Oliva et al 1994). Line ratios from ions in different stages of excitation along with a photoionization model can, thus, in principle be used to reconstruct the SED of the ionizing continuum. The tricky part is that the emission line flux does not only depend on the ionizing luminosity and EUV SED but also on the ionization parameter in the NLR (the ratio of ionizing flux to the local electron density). This makes the derived SED highly model-dependent on the radial and density structure of the NLR.

Following a first study of the ionizing continuum in the Circinus galaxy by Moorwood et al (1996a), Alexander et al (1999, 2000) have analyzed in detail three nearby Seyfert nuclei with this technique: Circinus (Seyfert 2 + starburst), NGC 4151 (Seyfert 1.5) and NGC 1068 (Seyfert 2). Starting with a (selected) compilation of ISO lines plus UV/optical/near-IR lines from the literature (typically 20 to 30 in each galaxy), Alexander et al find the best model from a scored fitting approach. In addition to the overall shape of the input SED (parameterized with 6 spectral points in the 10-500 eV range) the model depends on extinction, metallicity and the density structure/coverage of the gas. Alexander et al explore various NLR models and conclude that there are well defined, robust input SEDs for each of the three sources. These are shown in Fig.11. The Circinus data require a pronounced EUV bump peaking at about 70 eV and containing (in $\nu L_\nu$) at least 50% of the AGNs luminosity ($5 \times 10^9$ L$_\odot$, the other half is in the X-ray power-law, Moorwood et al 1996a). The black hole mass is less than about $4 \times 10^6$ M$_\odot$ (Maiolino et al 1998) so that the AGN's luminosity must be greater than 10% of the Eddington luminosity. The Circinus data thus are fully consistent with the standard AGN paradigm.

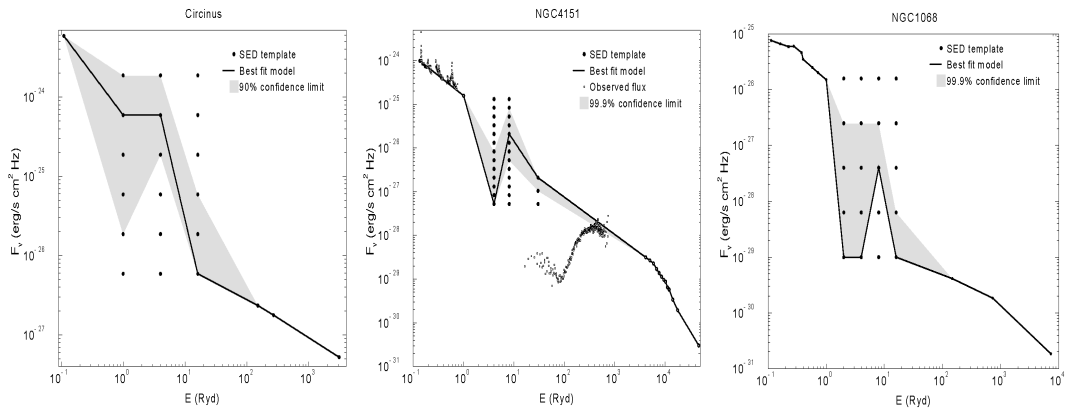

**Fig.11**

The derived SEDs for NGC 4151 and NGC 1068 appear to be different, possibly casting doubt on unified schemes (or the technique used). In both galaxies the SED turns down sharply just beyond the Lyman edge and then comes up again equally sharply for a peak around 100 eV to then finally drop down again and connect to the X-ray spectrum. Such a structured SED is not predicted by any accretion disk model, and it is not clear which physical mechanism might produce such a sharp emission feature at ~100 eV. Alexander et al propose as a solution that the NLR does not see the intrinsic SED of the AGN. The SEDs for both NGC 1068 and NGC 4151 can alternatively be interpreted by a EUV bump that has a deep absorption notch due to intervening absorption with a hydrogen column density of ~$4 \times 10^{19}$ cm$^{-2}$. Based on ISO SWS and HST observations Kraemer et al (1999) conclude as well that there is an absorber between the BLR and NLR of the very nearby Seyfert 1 galaxy NGC 4395. Further, there exists independent evidence for such intervening absorbing gas from UV observations in NGC 4151 (Kriss et al 1992, 1995). Such absorbers may be common in Seyferts (Kraemer et al 1998). These findings are thus consistent with the standard paradigm that all Seyfert galaxies are powered by thin accretion disks with a quasi-thermal EUV bump if in a (large) fraction of them the NLR sees a partially absorbed ionzing continuum.

Nevertheless caution needs to be exercized before one can fully trust these very encouraging new results. While Alexander et al explored a significant range of possible NLR geometries and density structures it remains to be shown whether the derived SEDs are robust if more complex models are considered. For instance, Binette et al (1997) conclude that the Circinus data can be fitted with a bump-less power law continuum if one allows for an component of optically thin (density bounded) gas clouds.

An additional interesting aspect of the IR coronal line work is that the NGC 4151 SWS line profiles show the same blue asymmetries that are characteristic of the optical emission line profiles in this and other Seyfert galaxies (Sturm et al 1999a). This result excludes the common interpretation – at least for NGC 4151 – that the profile asymmetries are caused by differential extinction in an outflow or inflow of clouds with a modest amount of mixed in dust. Sturm et al propose instead a geometrically thin but optically thick obscuring screen close to the nucleus of NGC 4151.

## 3.4 The Nature of Ultra-Luminous Infrared Galaxies

ULIRGs ($L(8-1000\mu m) \geq 10^{12}$ L , Soifer et al 1984, 1987) are mergers of gas rich disk galaxies (c.f. Sanders and Mirabel 1996, Moorwood 1996). Luminosities and space densities of ULIRGs in the local Universe are similar to those of QSOs (Soifer et al 1987, Sanders and Mirabel 1996). In a classical paper Sanders et al (1988a) have proposed that most ULIRGs are predominantly powered by dust enshrouded QSOs in the late phases of a merger. The final state of a ULIRG merger may be a large elliptical galaxy with a massive quiescent black hole at its center (Kormendy and Sanders 1992). Despite a host of observations during the last decade the central questions of what dominates on average the luminosity of ULIRGs and how they evolve are by no means answered. On the one hand their IR-, mm- and radio-characteristics are similar to those of starburst galaxies (Rieke et al 1985, Rowan-Robinson and Crawford 1989, Condon et al 1991b, Sopp & Alexander 1991, Rigopoulou et al 1996b, Goldader et al 1995,1997a,b, Acosta-Pulido et al 1996, Klaas et al 1999). Particularly compelling is the detection of a number of compact radio 'hypernovae' in each of the two nuclei of Arp 220 (Smith et al 1998). The extended optical emission line nebulae resemble the expanding 'superwind bubbles' of starburst galaxies (Armus et al 1990, Heckman et al 1990). On the other hand, a significant fraction of ULIRGs exhibits nuclear optical emission line spectra characteristic of Seyfert galaxies (Sanders et al 1988a, Armus et al 1990, Kim et al 1995, 1998, Veilleux et al 1995, 1997, 1999). Some contain compact central radio sources (Lonsdale et al 1993, 1995) and highly absorbed, hard X-ray sources (Mitsuda 1995, Brandt et al 1997, Kii et al 1997, Vignati et al 1999), all indicative of an active nucleus (AGN). With the advent of ISO sensitive mid- and far-IR spectroscopy have become available and have allowed a fresh look at the issues of the energetics, dynamics and evolution of ULIRGs.

*3.4.1 Signature for AGN vs. starburst activity: the ISO diagnostic diagram*

We have discussed in section 3.1.1 that AGN and starburst template galaxies have qualitatively very different spectral characteristics (Fig.4). In the 'ISO SWS diagnostic diagram' 13 bright ULIRGs studied with SWS and ISOPHOT-S (Genzel et al 1998) are located between pure AGNs and pure starbursts (Fig.5a). They appear to be composite objects but star formation dominates in most of them (especially when taking into account that the [OIV]/[NeII] line ratios are typically upper limits). In the simple 'mixing' model shown in Fig.5a, the average bright ULIRG in the study of Genzel et al (1998) has a ≤30% AGN contribution with ≥70% coming from star formation. In 4 of the 13 bright ULIRGs an energetically significant AGN is required by the measurements, either on the basis of a detection of strong high excitation lines (Mrk 273, NGC 6240), or on the basis of low UIB strength (Mrk 231, 05189-2524). In Mrk 231, 05189-2524 and Mrk 273 the AGN contribution may range between 40 and 80%.

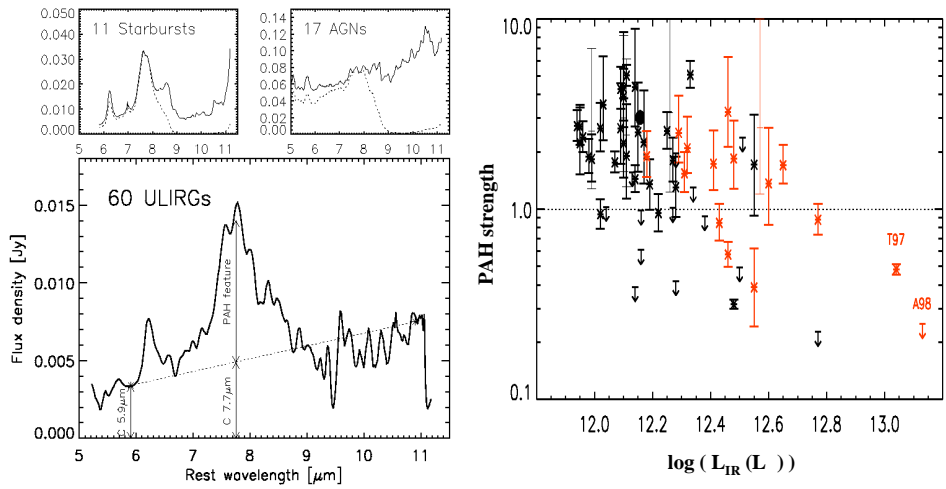

**Fig.12**

*3.4.2 ULIRGs as a class: the starburst-AGN connection*

Lutz et al (1998a), Rigopoulou et al (1999a) and Tran et al (1999) have extended this study to a larger sample by exploiting that the UIB strength criterion alone may be sufficient to distinguish between AGNs and starbursts (Fig.5a). The strong 6 to 11μm UIB features can be detected in much fainter and more distant sources (up to z=0.4 for the ISOCAM sample of Tran et al 1999). Note, however, that the UIB strength criterion rests heavily on the assumption that mid-IR features as well as continua are affected by the same extinction; otherwise a more detailed discussion is necessary (as in 3.3.1, see 3.4.3). The average spectrum of all 60 ULIRGs in the ISOPHOT-S sample is starburst-like (Fig.12, Lutz et al 1998). UIB strength and the ratio of mid-IR (5.9µm) and far-IR (60µm) flux densities are anticorrelated. 'Warm' ULIRGs ($S_{25}/S_{60} \geq 0.2$, ~20% of the sample) are AGN dominated and 'cold' ULIRGs are star formation dominated, in agreement with Sanders et al (1988b). The data are consistent with a model in which the UIB/FIR ratio is constant and mid-infrared extinction is one to a few magnitudes.

The UIB strength as a qualitative tracer of AGN vs. starburst nature can be exploited further to study the AGN-starburst connection as a function of luminosity and merger evolution. Fig.12 (right panel) shows the UIB strength as a function of infrared luminosity, for all ULIRGs of the ISOPHOT-S and ISOCAM samples, plus the two hyper-luminous ($L \sim 10^{13}$ L ) ULIRGs FSC15307+3252 (z=0.9, Aussel et al 1998) and P09104+4109 (z=0.44, Taniguchi et al 1997b). The AGN fraction appears to increase with luminosity, from ~20% for log L= 11.97-12.3, to ~50% for log L =12.3-13.2. Optical classifications lead to similar results. Kim et al (1998) classify as Seyfert 1/2s 26% of their ULIRGs for log $L_{IR}$=12-12.3, but 45% for log L=12.3-12.9. The ISO and optical data thus indicate that starbursts have an upper limit of $\leq 10^{13}$ L , consistent with the luminosity of a gas rich galaxy whose entire molecular gas mass is converted to radiation in a dynamical time and with the nucleosynthesis efficiency ε~0.001-0.007.

Lutz et al (1999a) find for 48 ULIRGs with ISOPHOT-S/ISOCAM data and good quality optical spectra that the optical and the ISO classifications of the individual galaxies agree very well if optical HII-galaxies and LINERs are both identified as starbursts. All but one of 23 ISO star formation dominated ULIRGs are HII/LINER galaxies. 11 of 16 ISO AGNs are Seyferts. All but one of the Seyfert 1/2s are ISO AGNs. Identification of infrared selected, optical LINERs as starbursts is quite plausible. LINER spectra in ULIRGs are likely caused by shock excitation in large scale superwinds and not by circum-nuclear gas photoionized by a central AGN (Heckman et al 1987, Armus et al 1990).

The evolutionary scenario proposed by Sanders et al (1988a) postulates that interaction and merging of the ULIRG parent galaxies triggers starburst activity which later subsides while the AGN increasingly dominates the luminosity and expels the obscuring dust. An implication of this scenario is that advanced mergers should, on average, be more AGN-like than in earlier stages when the interacting galaxies are still well separated. Rigopoulou et al (1999a) and Lutz et al (1998a) find, however, that powerful starbursts are present at the smallest nuclear separations. There is no obvious trend for AGNs to dominate with decreasing separation. This suggests that local and short-term factors, in addition to the global state of the merger, play a role in which of the energy sources dominates the total energy output. Relevant processes include the accretion rate onto the central massive black hole(s), the radiation efficiency of the accretion flow, and the compression of the interstellar gas on scales of a few tens of parsecs or less.

### 3.4.3 Hidden AGNs: the role of dust obscuration in ULIRGs

The active regions of ULIRGs are veiled by thick layers of dust. Fine structure and recombination line ratios imply equivalent 'screen' dust extinctions in ULIRGs (and LIRGs) between $A_V \sim 5$ and 50 (Genzel et al 1998). ISOPHOT-S and ground-based data confirm this evidence for high dust extinction (Lutz et al 1998a, Dudley&Wynn-Williams 1997, Dudley 1999). The most extreme case is Arp 220 where the ISO SWS data indicate $A_V(\text{screen}) \sim 50$ ($A_{25\mu m} \sim 1$), or an equivalent 'mixed model' extinction of 500 to 1000 mag in the V-band[8] (Sturm et al 1996). Smith et al (1989) and Soifer et al (1999) find similar values from the depth of the silicate feature and from mid-IR imaging. These large extinctions combined with a mixed extinction model solve the puzzle that starburst models based on near-IR/optical data cannot account for the far-IR luminosities ($A_V \leq 10$: Armus et al 1995, Goldader et al 1997a,b). If the emission line data are instead corrected for the (much larger) ISO-derived extinctions, the derived $L_{IR}/L_{Lyc}$ ratios are in reasonable agreement with starburst models (3.4.4).

Submm/mm CO and dust observations imply yet larger column densities than the mid-IR data ($\geq 10^{24}$ cm$^{-2}$, or $A_V \geq 500$, Rigopoulou et al 1996b, Solomon et al 1997, Scoville et al 1997, Downes and Solomon 1998). Is it possible, therefore, that most ULIRGs contain powerful central AGNs that are missed by the mid-IR data? Hard X-rays penetrate to column densities $\geq 10^{24}$ cm$^{-2}$. ASCA has observed a small sample of ULIRGs in the 2-10 keV band. About a dozen sources are common between ISO and ASCA (Brandt et al 1997, Kii et al 1997, Nakagawa et al 1999, Misaki et al 1999). In

---

[8] In the screen extinction model the dust is in a homogeneous screen in front of the source and the attenuation at $\lambda$ is given by $\exp(-\tau_d(\lambda))$ where $\tau_d(\lambda)$ is the optical depth of the dust screen at $\lambda$. An alternative, and probably more plausible scenario is that obscuring dust clouds and emitting HII regions are completely spatially mixed throughout an extended region. In that case the attenuation at $\lambda$ is $\tau_d(\lambda)/(1-\exp(-\tau_d(\lambda)))$ which changes much more slowly with wavelength.

Mrk 273, 05189-2524, NGC 6240 and 230605+0505 ASCA finds evidence for a hard X-ray source with $<L_X/L_{IR}> \geq 10^{-3}$. BeppoSAX observations show that the AGN in NGC 6240 is attenuated by a Compton thick ($N_H \sim 2 \times 10^{24}$ cm$^{-2}$) absorber (Vignati et al 1999). After correction for this absorption and depending on its filling factor the ratio of intrinsic AGN X-ray to IR luminosity is a 2 to $6 \times 10^{-2}$. In Mrk 231 a hard X-ray source is seen but is is weak ($<L_X/L_{IR}> \sim 10^{-3.5}$). ISO finds evidence for significant AGN activity in all of these sources as well. In sources classified by ISO as starburst dominated (Arp 220, UGC 5101, 17208-0014, 20551-4250, 23128-5919) ASCA also finds no hard X-ray source. The limit to the hard X-ray emission in Arp 220 corresponds to $\leq 10^{-4}$ of the infrared luminosity. For comparison, in Seyfert 1 and Seyfert 2 galaxies $<L_X/L_{IR}>$ is $10^{-1}$ and $10^{-2}$, respectively (Boller et al 1997, Awaki et al 1991). For radio quiet QSOs, the sample averaged SEDs of Sanders et al (1989) and Elvis et al (1994) give $L_X/L_{IR}=0.2$ and $L_X/L_{bol} \sim 0.05$. Although the statistics is still relatively poor at this point, the hard X-ray data do not present evidence for powerful AGNs that are completely missed by the mid-IR observations. Still, there are exceptions to this reasonable agreement between IR and X-ray data. The nearby galaxy NGC4945 fulfills all criteria of a pure starburst at optical to mid-IR wavelengths (Moorwood et al. 1996b, in Figure 5a NGC4945 is the starburst to the bottom right of Arp 220 and top left of M82). There is no evidence for an NLR or any other AGN indicator at these wavelengths. The [NeIII]/[NeII] line ratio is small and indicates that the starburst is aging (Spoon, priv.comm.). Yet ASCA and BeppoSAX data show that at its center lurks a powerful AGN, attenuated by a Compton thick foreground absorber (Iwasawa et al. 1993). As in NGC 6240 both the AGN (from the X-ray data) as well as the starburst (from the optical to IR data) can account for the entire bolometric luminosity of NGC4945. While NGC4945 is much less luminous than a ULIRG the case is puzzling and requires further study.

If mid-infrared continuum and UIB features suffer different obscurations, as in Seyfert 2 galaxies (3.3.1, Clavel et al 1998), the UIB strength criterion[5] looses its meaning. Instead it is necessary to directly compare the ratio of UIB-luminosity to total far-IR (60+100μm) luminosity even if this ratio depends on mid-infrared extinction. The UIB/FIR ratio in ULIRGs (Fig.10) is on average half of that in starbursts (and Seyferts). Following the discussion in 3.3.2 this suggests that at least half of the luminosity in the average ULIRG comes from star formation if the same UIB/FIR ratio holds as in other galaxies. Correction for extinction increases this fraction. If the average mid-IR extinction is $A_V$(screen)~15 and $A_{7.7}/A_V$~0.04, the UIB/FIR ratio is fully consistent with that in starburst galaxies.

Hence despite the large extinctions present in most ULIRGs optical/near-IR emission line diagnostics remain useful qualitative diagnostic tools in the majority of sources. For these objects mid-IR emission lines/UIB features can be used as quantitative tools for estimating the relative contributions of AGN and star formation. The reason is that optical and IR emission line diagnostics only rely on penetrating through dust in the disk to the NLR, and not through a high column density circum-nuclear torus to the central AGN itself. The large scale (≥100 pc) obscuring material is likely very patchy and arranged in thin (and self-gravitating) disks (as in Arp 220: Scoville et al 1998). Radiation and outflows from AGNs punch rapidly through the clumpy obscuring screen at least in certain directions.

*3.4.4 Properties and evolution of star formation in ULIRGs*

In the framework of starburst models with a Salpeter IMF and upper mass cutoffs of 50 to 100 M (Krabbe et al 1994, Leitherer & Heckman 1995) the ULIRG data are fit with burst ages/durations of t~$\Delta$t~5 to 100x10$^6$ years (Genzel et al 1998). In the case of NGC 6240 Tecza et al (1999) have directly determined the age of the most recent star formation activity (t=(2±0.5)x10$^7$ years) from the fact that the K-band light in both nuclei appears to be dominated by red supergiants. For that age and the measured (low) Br$\gamma$ equivalent width the duration of the burst $\Delta$t has to be about 5±2x10$^6$ years (Tecza et al 1999). Rigopoulou et al (1999a) find that there is no correlation between the gas content/mass of a ULIRG and the merger phase, as measured by the separation of their nuclei, in contrast to infrared galaxies of somewhat lower luminosities (LIRGs: Gao and Solomon 1999). Even very compact ULIRGs like Arp 220 and Mrk 231 still have a lot of molecular hydrogen.

Burst ages in ULIRGs thus are similar to those in other starbursts (3.2.2), are comparable to the dynamical time scale (between peri-approaches) and are much smaller than the gas exhaustion time (3.2.2) and overall merger age. There are likely several bursts during the merger evolution (3.2.1). A burst may be triggered as the result of the gravitational compression of the gas during the encounters of the nuclei (Mihos and Hernquist 1996). It is terminated within an O-star lifetime by the negative feedback of supernovae and super-winds. The models of Mihos and Hernquist suggest that the by far most powerful bursts occur when the nuclei finally merge and the circum-nuclear gas is compressed the most. The data indeed suggest that ULIRGs switch on only in the last ~20% of the merger history (Mihos 1999). The data do not, however, indicate that the most powerful starbursts in ULIRGs occur in the last phases with very small separations. Far-IR luminosities of ULIRGs do not change significantly with nuclear separation from separations of a few hundred parsecs to a few tens of kpc (Rigopoulou et al 1999a).

*3.4.5 Warm ULIRGs and QSOs*

'Warm' ULIRGs ($S_{25}/S_{60} \geq 0.2$) probably represent transition objects between ULIRGs and QSOs (Sanders et al 1988b). The ISO spectro-photometry discussed above strongly supports the conclusion of the earlier IRAS work that warm ULIRGs indeed contain powerful AGNs. For instance, the warm ULIRGs Mrk 231 (the most luminous ULIRG within $z \leq 0.1$) is very similar to the 'infrared-loud' but optically selected QSOs I Zw1 and Mrk 1014. Most of the warm ULIRGs have Seyfert spectra (Veilleux et al. 1997, 1999). High resolution HST and ground-based observations show that warm ULIRGs typically have a compact hot dust source at the nucleus, surrounded by a luminous host galaxy with embedded compact star clusters and large scale, tidal tails (Surace et al 1998, Surace & Sanders 1999, Lai et al 1998, Scoville et al 1999).

While the AGN is probably dominating, circum-nuclear star formation may nevertheless significantly contribute to the total luminosity of warm ULIRGs (Condon et al.1991b, Downes and Solomon 1998). In the case of Mrk 231 Taylor et al (1999) find a smooth, approximately circular radio continuum source of diameter ~1" (1 kpc), centered on the core-jet structure of the bright AGN. The extended radio source is associated with a massive rotating molecular disk seen at low inclination (Downes & Solomon 1998) and with a luminous ($2L_*$ for no extinction correction) near-IR stellar disk (Lai et al 1998). Applying the far-IR-radio correlation for star forming galaxies (Condon et al 1991, Condon 1992) to the parameters of Mrk 231's disk the implied far-IR luminosity is $1.3 \times 10^{12}$ L$_\odot$, or ~70% of the 60+100μm IRAS luminosity. If the red color of the disk's near-IR emission is due to extinction ($A_V$~12, Lai et al 1998) the corrected K-band luminosity of the disk is $L_K(0)^9$ ~$2 \times 10^{10}$ L$_\odot$. If the K-band emission comes from an old stellar population with $M/L_K$ ~100, the mass of the disk within the central kpc has to be $2 \times 10^{12}$ M$_\odot$ (an 8M$_*$ disk!). This mass is two orders of magnitude greater than the dynamical mass within the central kpc derived from the CO rotation curve (Downes and Solomon 1998). It is thus more likely that the K-band light comes from a population of (young) red supergiant or AGB stars. The required mass then is 20 to 100 times smaller and the ratio of bolometric to K-band luminosity is $\geq 10^2$ (Thatte et al. 1997). The red disk may then be responsible for a significant fraction of the far-IR luminosity of Mrk 231, in accordance with the radio data. The relative weakness of the mid-infrared line emission observed by ISO-SWS would then require very high extinction or, perhaps more likely, an aged starburst. The situations for the IR-loud QSOs IZw 1 and Mrk 1014 are similar but even more extreme stellar masses would be required there to account for the large K-band stellar luminosities in both galaxies (~$1.4 \times 10^{11}$ L$_\odot$) by an old stellar population. In both cases the dynamical masses implied by the molecular disk rotation excludes this possibility (Schinnerer et al 1998, Downes and Solomon 1998).

In summary of the ULIRG section, ISO has made substantial contributions to our understanding of ULIRGs. The resulting picture is fairly complex. There is no simple story to tell. ULIRGs do not seem to undergo an obvious metamorphosis from a starburst powered system of colliding galaxies to a buried and then finally a naked QSO at the end of the merger phase. Average luminosity and molecular gas content do not change with merger phase. AGN activity and starburst activity often occur at the same time. On average starburst activity does seem to dominate the bolometric luminosity in most objects. Yet the most luminous ULIRGs appear to be AGN dominated. 'Warm' ULIRGs are AGN dominated and 'cold' ULIRGs are star formation dominated. The relative contributions of starburst and AGN activity are in part controlled by local effects. Warm ULIRGs/infrared-loud QSOs are advanced transition objects with powerful central AGNs. In a few, well studied cases these AGNs are surrounded by luminous (aged?) starburst disks. Locally (U)LIRGs are a spectacular curiosity. They contribute only ~1% to the local infrared radiation field, and ~0.3% to the total bolometric emissivity (Sanders and Mirabel 1996, Heckman 1999). We will show in the next chapter that this fraction increases dramatically (factor $\geq 10^2$) at higher redshift.

---

[9] $L_K$ is here defined as the luminosity emitted in the K-band (in units of solar luminosities)

## 4. THE DISTANT UNIVERSE

In the local Universe surveyed by IRAS only 30% of the total energy output of galaxies emerges in the mid- and far-IR (Soifer & Neugebauer 1991). If the same were true at high redshifts as well, optical/UV observations alone would be sufficient for determining the cosmic star formation history. Spectacular (rest frame) UV observations of star forming galaxies at redshifts $\geq 3$ have become possible in the last few years (Steidel et al 1996, Lilly et al 1996). The inferred star formation rate per (comoving) volume element increases from z=0 by more than an order of magnitude to a peak at z~1-3 (Madau et al 1996, Pettini et al 1998, Steidel et al 1999).

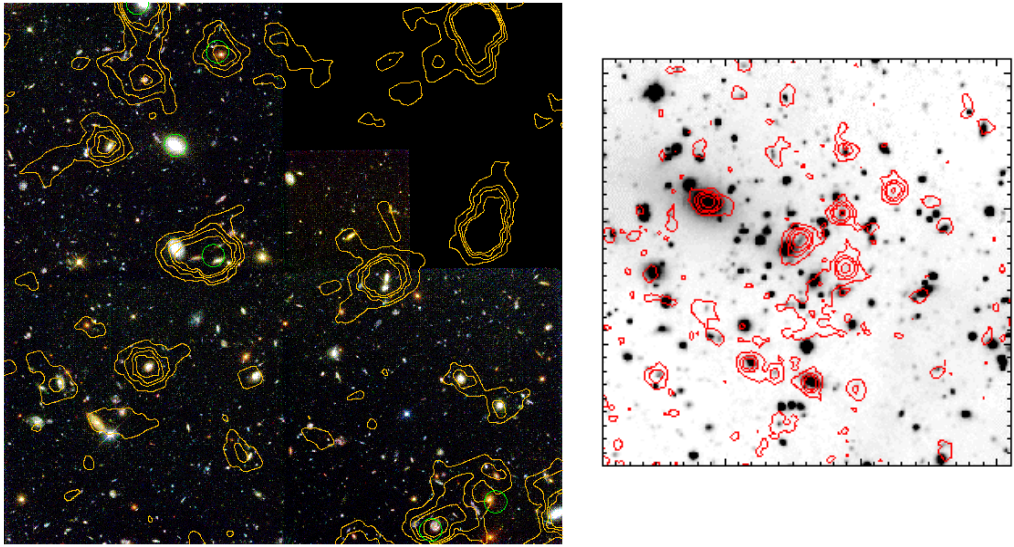

**Fig.13**

A significant new element in the picture of high-z star formation has emerged with the COBE detection of an extragalactic submillimeter background (Puget et al 1996). This diffuse background has an integrated intensity that is comparable to or larger than that of the integrated UV/optical light of galaxies (Dwek et al 1998, Hauser et al 1998, Lagache et al 1999). The implication is that there likely exists a very significant contribution of dust obscured star formation at high redshifts. The next major step clearly is to detect these high redshift infrared galaxies directly. With the exception of a few hyper-luminous and/or lensed objects (such as F10214+4724, Rowan-Robinson et al 1991) IRAS was only able to detect infrared galaxies to moderate redshifts (z~0.3). In the mid-infrared ISOCAM is ~$10^3$ times more sensitive and has 60 times higher spatial resolution than IRAS. In the far-IR with ISOPHOT the improvement in sensitivity is modest but the extension to longer wavelength (175µm) is of substantial benefit. Exploiting these improvements in deep surveys ISO has been able to provide for the first time a glimpse of the infrared emission of galaxies at z≥0.5.

## 4.1 Mid-infrared surveys

The various ISOCAM surveys trade off depth and area (see Rowan-Robinson et al 1999 for a compilation). For high redshift work the most meaningful are the 15µm (LW3 filter) surveys. At 6.75µm (LW2) the stellar contamination is very high and a careful screening at other wavelengths is necessary. Various strategies have been successfully developed to overcome the cosmic ray induced glitches that mimic faint sources (Starck et al 1999, Desert et al 1999, Serjeant et al 1999). At the time of writing this article the published work reports 15µm detections of ~1400 galaxies, with flux densities between ≤30µJy and ≥0.3 Jy. Shallow, deep and ultra deep surveys were performed in the Lockman Hole and the Marano field (IGTES surveys, Cesarsky and Elbaz 1996, Elbaz et al 1999a). They are complemented on the faint end by the surveys made on deep IR fields centered on the southern and northern Hubble Deep Fields (Fig.13) as observed by Rowan-Robinson and colleagues (HDF(N): Serjeant et al 1997, Aussel et al 1999a {Fig.13}, Desert et al 1999; HDF(S): Oliver et al 1999, Elbaz et al 1999a). Yet deeper images are available from lens magnification in the direction of distant clusters (Metcalfe et al 1999, Altieri et al 1999: Fig.13, Barvainis et al 1999). Distant clusters also are surveyed to compare the evolution of galaxies in different environments (Lemonon et al 1998, Fadda and Elbaz 1999). The bright end of the luminosity function is explored by the largest area survey, the 12 square degree

ELAIS program (Rowan-Robinson et al 1999, Serjeant et al 1999). Levine et al (1998) have reported the first results of a program to observe 500 candidate extragalactic sources from the IRAS faint source survey with ISOCAM and ISOPHOT (Levine et al 1998). This ISO-IRAS faint galaxy survey efficiently picks up moderate redshift (z=0.1-0.4) (U)LIRGs.

*4.1.1 Source counts: evidence for strong evolution*

Fig.14 (from Serjeant, priv.comm. and Elbaz et al 1999a,b) gives the integrated and differential source counts derived from these surveys. Given the difficulties of reliable faint source detection and photometry mentioned above it is reassuring that the counts of the different surveys agree quite well. Starting with IRAS and going to fainter flux densities the number counts initially lie on a linear (log-log plot) extrapolation of the IRAS counts and do not require strong evolution. However, from >10mJy down to 0.4mJy the counts increase rapidly with a slope in the integrated counts of $\alpha=-3$. This is significantly steeper than expected in a Euclidean model without luminosity/density evolution ($\alpha=-2.5$). Below 0.4mJy down to the faintest flux densities sampled the slope is flatter ($\alpha=-1.6$) and the counts appear to converge. In the differential source counts (right inset of Fig.14, Elbaz et al 1999a, 2000) this leads to a prominent hump peaking at ~0.4mJy. At the peak of the hump the observed source counts are an order of magnitude above non-evolution models, obtained by extrapolating the local IRAS 12$\mu$m luminosity function (Rush et al 1993, Fang et al 1998). The mid-infrared source counts require strong cosmic evolution of the mid-infared emission of galaxies (Oliver et al 1997, Roche & Eales 1999, Clements et al 1999, Elbaz et al 1999a,b, Serjeant et al 1999).

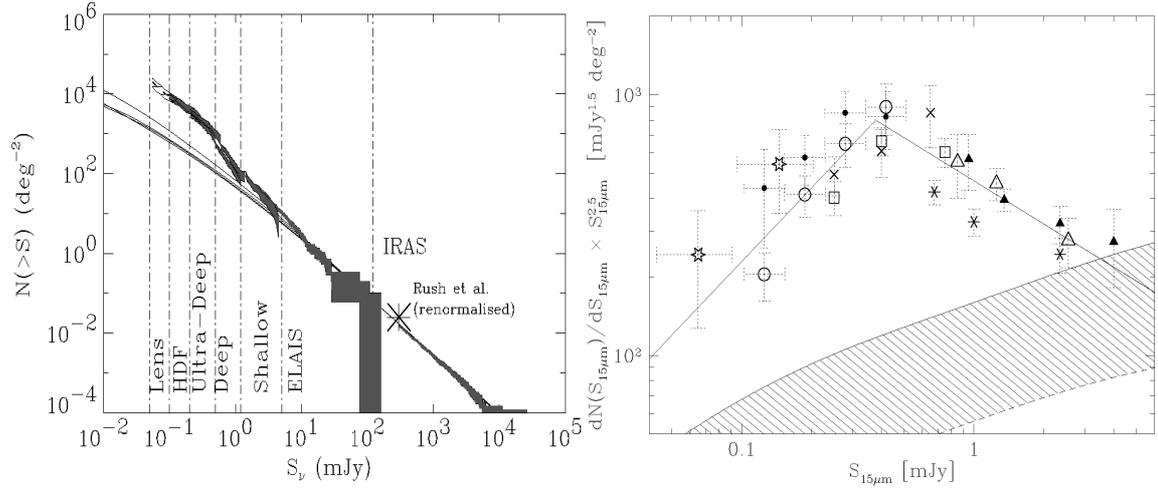

**Fig.14**

The 15μm integral number counts agree with model predictions over more than five orders of magnitude in flux density (Franceschini et al 1994, 2000, Pearson & Rowan-Robinson 1996, Guiderdoni et al 1997, 1998, Xu et al 1998, Roche & Eales 1999). Common to all these models is the assumption of strong luminosity and/or density evolution of dusty star formation in bright normal spirals and starburst galaxies, with varying contributions from AGNs. We discuss now in more detail the models by Xu et al (1998), since they are based on the most recent ISO SWS-ISOPHOT-S spectra of template galaxies to determine accurate, mid-IR k-corrections. These are important because of the structure caused by UIB features in the 6-13μm spectra of most galaxies. The models of Pearson, Rowan-Robinson, Franceschini, Roche & Eales and Guiderdoni et al lead to similar results. Xu et al construct a local luminosity function from IRAS observations (Shupe et al 1998) and assign to each of these galaxies a mid-IR spectrum composed of three categories (cirrus/PDR, starburst, AGN, or a mixture). They then extrapolate the local luminosity function to higher redshift by applying the appropriate k-corrections and evolution. The predicted differential, Euclidean normalized source counts for the LW3 ISOCAM filter show distinct differences between luminosity evolution (L(z) $\propto$ $(1+z)^3$) and density evolution ($\rho$(z) $\propto$ $(1+z)^4$). For pure density evolution the predicted source counts are fairly flat from 1 Jy to 100µJy. For luminosity evolution, however, a characteristic bump is predicted at around 300µJy, with a sharp falloff of the counts at fainter flux densities. This is similar to what is actually observed by Elbaz et al (1999a, Fig.14) although the amplitude of the hump in the differential source counts requires an even steeper evolution than considered by Xu et al (e.g. a combination of $(1+z)^3$ luminosity evolution and $(1+z)^6$ density evolution, for the starburst component only, at z<0.8-0.9: Franceschini et al 2000). The interpretation of the bump is that with luminosity evolution and flat or even 'negative' k-correction (between z=0.5 and 1, Elbaz et al 1999b) the number counts increase initially with decreasing flux density much faster than Euclidean. The unfavorable k-correction and the decreasing slope of the available comoving volume cause the number counts to drop rapidly with increasing z for z$\geq$1. For sources near the bump, the luminosity evolution model predicts a broad redshift distribution, with a median at z~0.9 but extending with significant probability to redshifts >2. In contrast the pure density evolution model predicts a much smaller median redshift (~0.5) and essentially no sources with z$\geq$1. When the mid-infrared counts are compared to K-band counts at the same energy ($\nu S_\nu$) the mid-infrared sources contribute only about 10% of the energy at the bright end but more than 50% at $S_{15\mu m}$~a few hundred µJy (Elbaz et al 1999a). The ISOCAM sources thus must represent an important sub-class of the optical/near-IR galaxies that are dusty and infrared active.

The situation is less clear at 6.75μm (Goldschmidt et al 1997, Taniguchi et al 1997a, Aussel et al 1999a,b, Sato et al 1999, Serjeant et al 1999, Flores et al.1999a). The stellar contamination is larger. The local luminosity function and template spectra appropriate for the galaxies are less well known than at 15μm. The number of confidently detected sources is also much smaller. They are (more) compatible with a non-evolution (i.e. passive evolution) model with a significant contribution of ellipticals/S0s (Roche & Eales 1999). They are also consistent with the Pearson & Rowan-Robinson (1996) luminosity evolution model but not with that of Franceschini et al (1994) (Serjeant et al 1999).

*4.1.2 Nature of ISOCAM sources: luminous star forming galaxies at z~0.7*
At the time of writing this article, the properties of the ISOCAM galaxies at other wavelength bands have been studied in detail in two fields: HDF (N) (Mann et al. 1997, Rowan-Robinson et al 1997, Aussel et al 1999a) and the 1415+52 field of the Canada-France Redshift Survey (CFRS: Flores et al 1999a,b). In HDF(N) Aussel et al (1999a,b) have extracted a list of 38 galaxies (≥100 μJy, ≥99% confidence) with optical identifications in the catalog of Barger et al (1999). Of these 26 galaxies have known redshifts. In the $(10')^2$ CFRS 1415+52 field Flores et al (1999b) detect 78 significant (≥3σ) 15μm sources (≥250μJy), 22 of which have spectra and redshifts. The median redshift in both fields is near 0.7 and most redshifts range between 0.4 and 1.3. The ISOCAM galaxies have the optical colors of Sbc-Scd galaxies. The majority of the galaxies are disk or interacting systems, the remainder are irregulars and E/S0s. The one elliptical in the HDF(N) sample is assoicated with an AGN (z=0.96). Typical absolute K-band magnitudes HDF(N) range between –22 and –25, similar to but on average somewhat fainter than an $L_*$ galaxy (M(K)=-24). The ISOCAM galaxies thus are mainly luminous disk/interacting galaxies and are definitely not part of the faint blue galaxy population responsible for the excess in faint B-band counts (Ellis 1997).

The local luminosity function at the mid-IR luminosities characteristic of the ISOCAM galaxies (~$5 \times 10^{10}$ L ) is dominated by AGNs (Fang et al 1998). However, the rest-wavelength B-band spectroscopy in both the HDF(N) and CFRS fields (Flores et al 1999b, Aussel et al 1999b) and the rest-wavelength R-band observations of the HDF(S) field (Rigopoulou et al 1999b) suggest that most of the ISOCAM galaxies are dominated by star formation, with no more than ~1/3 being AGNs. The majority of the galaxies in the CFRS field (15 or 70%) have 'e(a)' optical spectra[10] characteristic of 'post-starburst' systems (age ≥500 Myrs), or active starburst galaxies galaxies with large, differential intrinsic dust extinction. The latter explanation is the almost certainly the correct one (Aussel et al 1999b, Rigopoulou et al 1999b). For 19 galaxies with complete radio to UV SEDs, Flores et al (1999b) identify more than half as (highly reddened) starbursts. Dusty starburst galaxies such as M82 (Kennicutt 1992), many bright ULIRGs (Liu and Kennicutt 1995) and 50% of the (U)LIRG sample studied by Wu et al (1998) have e(a) spectra in the B-band. Local e(a) galaxies have large Hα equivalent widths, at the same time demonstrating active current star formation and differential dust extinction. Rigopoulou et al (1999b) have recently carried out near-IR, VLT-ISAAC spectroscopy of a sample of $0.6 \leq z \leq 1.3$ ISOCAM galaxies in the HDF(S) field. The ISOCAM HDF(S) galaxies have large (50-100 Å) Hα equivalent widths, providing compelling evidence that most of them are active starbursts. The simultaneous presence of heavily dust enshrouded present star formation and less extincted older star forming activity probably indicates several starburst episodes (3.2.1, 3.4.4).

---

[10] 'e(a)' (Poggianti and Wu 1999) or 'S+A' (Hammer et al. 1997) galaxies have moderate EW ([OII]) line emission and at the same time strong Balmer (Hδ etc.) absorption or (at low spectral reolution) a large 3550-3850 Å Balmer continuum break, characteristic either of A-stars or very large extinction. In the A-star model the characteristic age of the (optically visible) star formation is ~0.5 -1 Gyrs. Galaxies with Balmer absorption but no emission lines are called 'E+A', or 'k+A'.

The infrared derived star formation rates are substantially greater than those determined from the [OII] (or Hα) lines. On the basis of the SEDs, Flores et al find that the median 8-1000μm luminosity of the CFRS sample is $\sim 3 \times 10^{11}$ L$_\odot$ (star formation rate ~30 to 50 M$_\odot$ yr$^{-1}$). The HDF(N) results are similar. Rigopoulou et al and Franceschini (priv.comm.) deduce star formation rates of a few tens of M$_\odot$ yr$^{-1}$ from the Hα emission but typically three times greater values from the mid-IR data. Most of the faint ISOCAM galaxies thus appear to be LIRGs. A smaller fraction (~25%) of the CFRS sources have ULIRG-like luminosities ($\geq 10^{12}$L$_\odot$), in agreement with the work of Rowan-Robinson et al (1997) for HDF(N). Other confirmations of the identification of the faint ISOCAM galaxies as luminous starbursts come from observations of several galaxies near z~1 that are lensed by foreground clusters (Lemonon et al 1998, Barvainis et al 1999). All e(a) galaxies in the z=0.2 cluster A1689 are 15μm ISOCAM sources (Duc et al 2000). Of the 'k+A' and e(a) galaxies in Coma only those with emission lines have excess mid-IR emission indicative of active star formation (Quillen et al 1999).

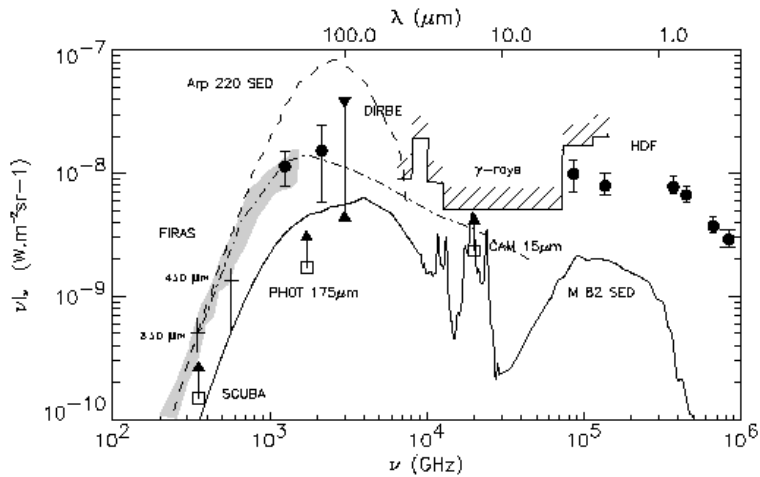

**Fig.15**

In the LW2 filter the situation is similar but the relative number of AGNs and ellipticals is proportionally larger. Flores et al (1999a) identify 40% of the 15 CFRS/6.75μm galaxies with spectra as AGNs, 53% are active starbursts or S+A-galaxies. Aussel et al (1999b) classify 4 of the 6 confidently detected 6.75μm sources in HDF(N) as elliptical galaxies.

For the great majority of ISOCAM sources the far-IR counterparts are not yet known. Still template spectra and the overall background constraint can be used to draw first order conclusions (Fig.15). If all the ISOCAM sources had ULIRG spectra as cool as Arp 220 (L(80μm)/L(8μm)~70[11]), they would significantly overproduce the far-IR/submm background and would give rise to counts at 850/175 μm well in excess of the observations (Aussel et al 1999b, Elbaz et al 2000). Thus, on average, they must have SEDs at least as warm as LIRGs (like M82: L(80µm)/L(8µm)~5..12), in which case they still produce 30 to 60 % of the far-IR background (Fig.15). The contribution of the ISOCAM sources to the background would only be small if they were predominantly dust enshrouded AGNs (Fig.15). This interpretation is not supported, however, by the HDF(N,S) and CFRS samples.

---

[11] Arp 220 has an extremely cool SED. Average ULIRGs have L(80μm)/L(8μm)~25 which would still exceed the far-IR/submm background but only marginally so, yet leaving no space for a higher z population.

## 4.2 Far-infrared surveys

The SED of actively star forming galaxies peaks at $\lambda$~60-80µm. For observations at $\lambda \geq 200$µm the 'negative' k-correction is advantageous for detecting high-redshift galaxies and at the same time, such measurements give the total luminosity directly, without a model-dependent bolometric correction. For these reasons, several surveys were undertaken with the ISOPHOT at 175µm. A one square degree field in the Lockman hole was surveyed as part of the Japanese ISO cosmology project (Kawara et al 1998, 1999, Fig.16). Two fields in the southern Marano area and two fields in the northern ELAIS fields with a combined area of 4 square degrees constitute the FIRBACK program (Puget et al 1999, Dole et al 1999). An additional 'serendipity' survey used the slews between pointed observations to survey the sky at ~200µm at the ≥2 Jy level (Stickel et al 1998, 1999). This survey extends IRAS far-IR SEDs to longer wavelengths for a few thousand nearby galaxies. The depth of the two deep surveys is ~100 mJy (5$\sigma$). The key problems for the far-IR detection of faint individual sources are detector noise and large, spatially varying backgrounds: Galactic cirrus, extragalactic backgrounds and (to a lesser extent) zodiacal dust emission. To minimize the impact of cirrus the deep survey areas were chosen to lie in regions known to have low HI-column densities and IRAS 100μm surface brightness. In the best areas the surveys are limited by the fluctuations in the extragalactic background (Lagache & Puget 1999). In the case of the Lockman hole and ELAIS regions a comparison with a second map at 90μm gives additional useful information. In the final source lists only detections with flux densities ≥120 mJy were taken into account. At that level reliability and completeness are good and far above the expected fluctuations in the cirrus and extragalactic backgrounds. The sources detected therefore are likely individual galaxies. At the 100mJy cutoff level Kawara et al report 45 175μm sources (and 36 95μm sources) in the Lockman field and Dole et al find 208 175μm sources in the combined Marano and ELAIS fields. In a 0.25 square degree Marano field Puget et al detect 22 sources. Fig.16 shows the counts from all three references, in comparison to models A and E of Guiderdoni et al (1998). At the lowest flux densities the number counts are a factor 4 to 10 above the predictions of non-evolving models. The counts are consistent with those that include strong evolution as well as a significant population of $z \geq 1$ ULIRGs (e.g. model E of Guiderdoni et al 1998).

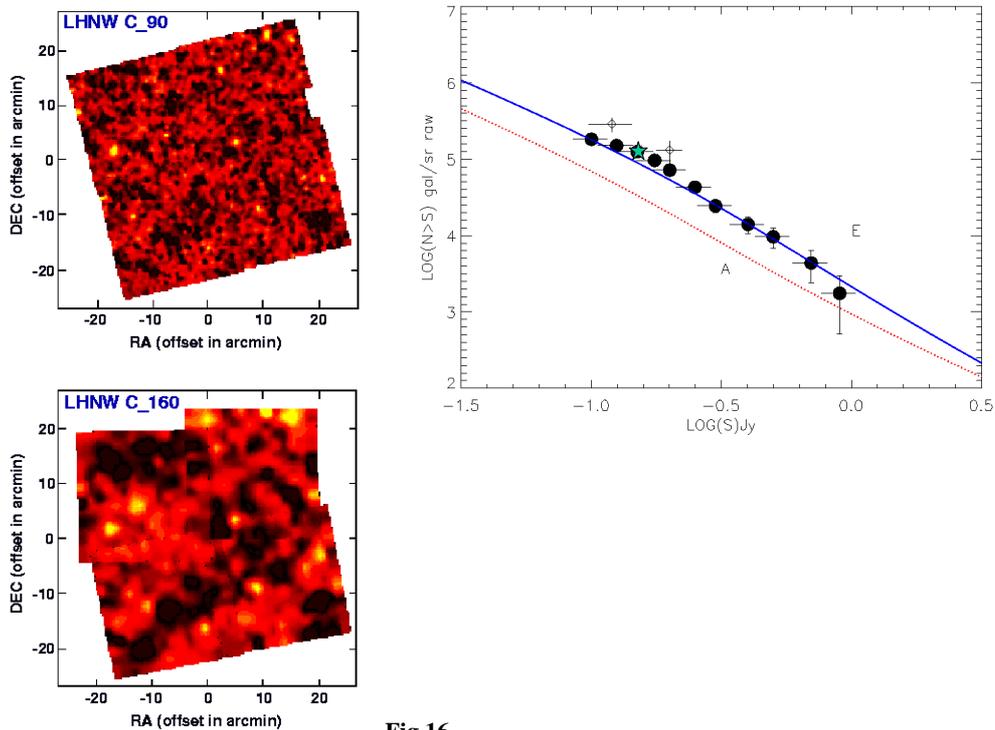

**Fig.16**

The integrated counts are an order of magnitude greater than the 100µm IRAS counts at the same flux level. In the ELAIS area 31 sources have a 90µm counterpart. The average 175/90 flux density ratio is $\geq 2$, as compared to ~1 for a sample of local sources (Stickel et al 1998). One part of the Marano ISOPHOT field overlaps with the deep ISOCAM survey area. At most 50% of the ISOPHOT sources have 15µm counterparts (Elbaz et al 2000). For those the 175µm/15µm flux density ratios correspond to what one would expect for relatively nearby (z~0.5) galaxies with Arp 220-like spectra. Scott et al (1999) have observed 10 FIRBACK sources with SCUBA. The average 850 (and 450) µm/175µm flux density ratios also favor a low mean redshift (~0.3) although the degenerate dust temperature/ redshift relation would also permit in principle solutions with z up to 2. The far-IR sources not detected by ISOCAM could be at higher redshift. The interplay of the k- corrections is such that ISOCAM cannot see galaxies with an Arp 220 SED located at z~2.

The integrated counts constitute less than 10% of the COBE far-IR background (Hauser et al 1998). This suggests that ISOPHOT is detecting the tip of the iceberg of a new population of very luminous, moderate to high redshift galaxies. Most of the background is in sources with 175µm flux densities between a few and 100mJy.

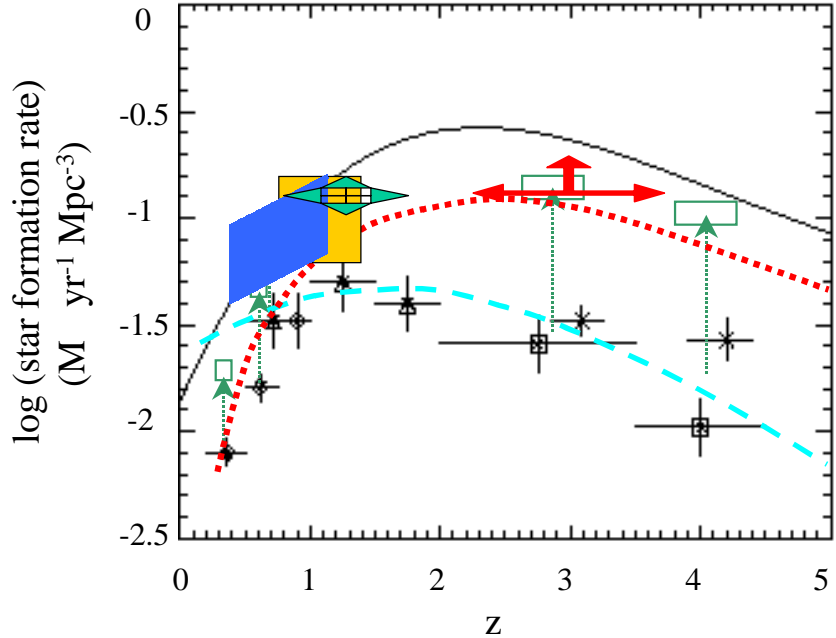

**Fig.17**

## 4.3 Star formation at high redshift: a synthesis

From the survey of the 1415+52 CFRS field Flores et al (1999b) conclude that 60±25% of the star formation at z≤1 is associated with infrared emission. Including the dusty luminous starbursts detected by ISOCAM the total star formation rate (per unit comoving volume) at z≤1 is 2.9±1.3 times as large as that deduced from the CFRS restframe UV observations (without correction for extinction). Rowan-Robinson et al (1997) found an even larger correction factor from an analysis of the ISOCAM HDF(N) data. If the ISOPHOT 175µm sources are ultra-luminous starburst galaxies at z~1 the far-IR data require a star formation rate also ~3 times larger than predicted by UV observations (not extinction corrected) interpolated into this redshift regime. Finally, the 850µm source detections with the SCUBA bolometer camera require a star formation rate at z~3 that is ~3.5 times greater than that derived from Lyman-dropout galaxies (Hughes et al 1998, Barger et al 1998, Ivison et al 1998). This assumes that the 850µm sources are powered by star formation. To explain the entire COBE far-IR/sub-mm background requires the star formation rate at z≥1 to be greater by another factor ≥2 (Lagache et al 1999b). The recent Hα-based star formation rate at z~1.3 (Yan et al 1999), on the other hand, is in good agreement with the far-IR/submm estimates. The large discrepancy between the UV and IR/submm estimates disappears if the star formation rates estimated from the Lyman-dropout galaxies are corrected for extinction (Pettini et al 1998, Meurer et al 1999, Steidel et al 1999). The data are summarized in Fig.17 (the 'Madau' plot: adapted from Lagache et al 1999b, Steidel et al 1999).

The good agreement between the UV/optical and IR/submm star formation histories shown in Fig.17 may be a lucky coincidence, however. The IR/sub-mm and UV/optical measurements trace different source populations. At the level expected from the extinction corrected, UV star formation rates Chapman et al (1999) have detected at 850µm only one of about a dozen Lyman break galaxies. The reddening of the ISOCAM galaxies is only slighty higher than average (Elbaz et al 2000), and thus their true output cannot be correctly derived from optical and UV data alone. The extinction corrections in the UV/optical are poorly known. In the case of the far-IR/submm sources the redshifts are uncertain, and the contribution of AGNs is unknown. Based on the X-ray background Almaini et al (1999) set a limit of ~20% to the AGN contribution to the far-IR/submm background. Only 10 to 40% of the far-IR/submm background are resolved into individual sources. The derivation of star formation histories from the overall far-IR/submm background crucially depends on the (uncertain) source SEDs used.

Nevertheless a reasonably consistent picture seems to emerge. The cosmic star formation rate steeply increases between the current epoch and z~1 and stays flat from z~1 to at least 4. The 'quiescent' mode of disk star formation (with typical gas exhaustion time scales of several Gyr) cannot explain the z≥1 data. Such a scenario (model 'A' of Guiderdoni et al 1998, dashed in Fig.17) can be definitely excluded by the extinction corrected UV points, the mid-infrared points and the far-IR/submm points. A rapidly evolving 'burst' mode (exhaust time scales a few hundred Myrs) is required, probably associated with the much increased merger rate at high redshift ( $(1+z)^\delta$, $\delta=2..6$, Zepf and Koo 1989, Carlberg 1990, Abraham et al 1996). In the semi-analytic models of Guiderdoni et al (1998) the fraction of this burst mode in the numbers of stars formed increases with $(1+z)^5$ and dominates the cosmic star formation at z≥1. To also fit the far-IR/submm data the fraction of violent mergers within the starburst population leading to ultra-luminous starbursts must rapidly increase with redshift (Guiderdoni et al 1998, Blain et al 1999). In the model 'E' of Guiderdoni et al the fraction of star formation in ULIRGs increases from 8% at z=1 to 27% at z=3. Blain et al (1999) introduce an 'activity parameter' of mergers, defined as the inverse of the product of the fraction of mergers leading to luminous starbursts/AGNs $f_b$ and the duration of an activity event $\Delta t_b$. In the Blain et al models the activity parameter must increase by about a factor of 3 from the present day to z~0.7 (the '15μm (ISOCAM) epoch'), by ~10 to z~1 (the 'ISOPHOT epoch' ? ) and by ~160 to z~3 (the 'SCUBA epoch'). Hence the relative contribution of ULIRGs to the luminosity function increases by a factor of several hundred between the local

Universe and the peak of the cosmic star forming activity at z~2-3. The models of Blain et al and Guiderdoni et al thus are quite consistent with each other and with the observations. The starburst galaxies sampled by ISOCAM have characteristic ages less than a few hundred Myrs (Flores et al 1999b, Aussel et al 1999b), at least an order of magnitude smaller than the gas exhaustion time scale in present day quiescent disks. The characteristic ages of nearby ULIRGs (100 $L_*$) that may be representative of the 175µm and 850µm populations are $\geq$10 Myr (Genzel et al 1998, Tezca et al 1999) and they probably make up only a small fraction of all merger events. The ultra-luminous mergers may be related to the formation of large elliptical galaxies and bulges (Kormendy and Sanders 1992). As a result of these powerful mergers the cosmic star formation rate may stay flat or even increase to redshifts $\geq$4 (Blain et al 1999, Pei and Fall 1995).

# 5. CONCLUSIONS AND OUTLOOK

*Normal galaxies:* ISO has provided progress on many fronts. The impulsive heating mechanism of single UV (and optical?) photons impinging on very small dust grains/large molecules and emitting in the near-/mid-IR appears now firmly established. The mid-IR SED is a valuable diagnostic of the activity level in galaxies. The mid-IR emission from early type galaxies is composed of stellar, interstellar and AGN contributions which can now be separated. In nearby spirals the far-IR data require cool, equilibrium dust with a near Galactic, dust-to-gas ratio. ISO has added tantalizing results on the nature of the halos and outer disks of spiral galaxies but has not solved the overall puzzle of the dark matter residing there. Deeper images are required and should become available with SIRTF, NGST and ground-based telescopes.

*Starburst galaxies (and the starburst/AGN connection):* ISO has added infrared spectroscopy and spectro-photometry to the astronomer's arsenal of powerful analytical tools. Dust enshrouded AGNs and starburst galaxies can be qualitatively and quantitatively distinguished. ISO has confirmed that infrared luminous starburst galaxies typically have low nebular excitation. The likely explanation is that starbursts are episodic and rapidly aging. Star formation provides strong negative feedback that tends to quench active starbursts in little more than an O star's lifetime.

*AGNs:* The ISO data generally support unified accretion disk+circum-nuclear torus/disk models. Emission from AGN-heated, thick tori or warped disks dominate the mid-IR emission of QSOs, radio galaxies and Seyfert nuclei. Based on a combination of mid-IR spectro-photometry, radio/IR imaging and radiative transport models, star formation likely plays an important role in accounting for the far-IR emission of many Seyfert galaxies. However, the issue of what powers the (thermal) far-IR emission of QSOs and radio galaxies remains uncertain. Mid- and far-IR spectroscopy of powerful AGNs with SIRTF and SOFIA will provide the next major step in this field. A promising new method has emerged to re-construct the (hidden) EUV SEDs of AGNs from optical/infrared 'coronal' line ratios. SIRTF will tell how useful this tool is for quantitative studies of different classes of AGNs.

*(U)LIRGs:* With ISO much progress has been made in understanding the properties, energy sources and evolution of (ultra-) luminous infrared galaxies which are a spectacular curiosity locally, but very common and important in the early Universe. (U)LIRGs are composite objects containing both powerful AGNs and starbursts. Star formation plausibly dominates the luminosity of most sources. At the highest luminosities AGNs take over. The standard paradigm of ULIRGs as a (late) phase in the merger of two gas rich galaxies is confirmed but their evolution is more complex than considered before. ULIRGs do not seem to undergo an obvious metamorphosis from a starburst powered system of colliding galaxies to a buried and then finally a naked QSO at the end of the merger phase. Local effects play an important role. SIRTF will increase the statistics and quality of the mid-IR spectroscopy, and will also be able to apply the new IR spectroscopic tools to higher redshift sources. XMM will do the same at hard X-rays. SOFIA promises sensitive far-IR spectra to tackle the problem of the weak far-IR line emission seen by ISO in nearby UILRGs.

*High-z, dusty star formation:* COBE, ISO and SCUBA have clearly demonstrated that dusty star formation plays an important role in the early Universe. This is a very exciting development. Yet the present data have barely scratched the surface of this important emerging field and their interpretation is still in a stage of infancy. New models and follow-up work are in progress and updates give an increasingly detailed picture. ISOCAM has identified an important population of active starburst galaxies (mainly LIRGs) at $z<1.5$. This population accounts for 10 to 60 % of the far-IR/submm background. ISOPHOT has begun to detect a ULIRG component at moderate redshift ($z\sim0.2$ to $>1$) that may account for an additional 10 % of the background. Submm observations (SCUBA etc.) are detecting a component of high redshift (z≥2) ULIRGs which are making up the remainder of the background. Deeper surveys with SIRTF, SOFIA, FIRST and ALMA will be one prime objective of the next decade. Identifying counterparts at other wavelengths and studying their nature will be another. The surveys will resolve most of the far-IR/submm background into individual sources. IR spectroscopic follow-up (with ground-based telescopes, SIRTF, FIRST) and hard X-ray observations (Chandra, XMM) will establish the redshifts and will determine what fraction of the IR sources are active starbursts and what fraction are dusty Seyferts/QSOs. We can look forward to an exciting period!

*Acknowledgements.* *We would like to thank the many colleagues who have sent us re- and preprints, material prior to publication or given us inputs, comments and advice, and in particular P.Andreani, H.Aussel, R.Barvainis, J.Clavel, P.Cox, H.Dole, D.Elbaz, J.Fischer, A.Franceschini, M.Haas, G.Helou, D.Hollenbach, M.Kaufman, O.Laurent, D.Lemke, S.Lord, M.Luhman, D.Lutz, R.Maiolino, S.Malhotra, H.Matsuhara, F.Mirabel, A.F.M. Moorwood, T.Nakagawa, H.Okuda, J.-L. Puget, D.Rigopoulou, B.Schulz, S.Serjeant, B.Smith, M.Stickel, E.Sturm, Y.Taniguchi, M.Thornley and D.Tran. We would like to thank Susanne Harai, Dieter Lutz, Linda Tacconi and Lowell Tacconi-Garman for help with the manuscript and for useful comments. We are especially grateful to F.Mirabel and A.F.M. Moorwood for their comments on the manuscript.*

# FIGURE CAPTIONS

**Figure 1.** ISOCAM/IRAS color diagram for diiferent types of galaxies. The decrease in LW2/LW3 ratio for 60/100 flux ratios >0.5 (AGN/starbursts and dwarf galaxies) can be explained by a combination of the destruction of the PAH features (LW2 band) in the intense radiation field and increased emission from (very small) warm grains (LW3) (from Vigroux et al 1999).

**Figure 2.** ISOPHOT 175μm map of M31 (north up, east to the left).The emission is dominated by a 10 kpc ring, with numerous bright condensations and a fainter ring at 14kpc (from Haas et al 1998b). The ratio of ring to nuclear emission is much greater than in the IRAS bands (Habing et al 1984).

**Figure 3.** Combined LWS+SWS spectra of galaxies (> 6 octaves). Left: Combined SWS/LWS spectrum of the Circinus galaxy (Moorwood 1999, Sturm et al 1999b). The $H_2$ lines and low excitation atomic/ionic fine structure lines ([FeII],[SiII], [OI], [CII]) sample photodissociation regions (PDRs[4]: Sternberg and Dalgarno 1995, Hollenbach & Tielens 1997), shocks (Draine et al 1983, Hollenbach & McKee 1989), or X-ray excited gas (Maloney et al 1996). Hydrogen recombination lines and low lying ionic fine structure lines (excitation potential <50 eV: [ArIII], [NeII], [NeIII], [SIII], [OIII], [NII]) sample mainly HII regions photoionized by OB stars (Spinoglio & Malkan 1992, Voit 1992), although ionizing shocks may contribute in some sources (e.g. Contini & Viegas 1992, Sutherland et al 1993). Ionic lines from species with excitation potentials up to ~300 eV (e.g. [OIV],[NeV], [NeVI], [SiIX]) probe highly ionized, 'coronal' gas and require very hard radiation fields (such as the accretion

disks of AGNs), or fast ionizing shocks. Line ratios give information about the physical characteristics of the emitting gas. Extinction corrections are small ($A(\lambda)/A(V)$~0.1 to 0.01 in the 2 to 40μm region). Right: The starburst galaxy M82 (top): low excitation lines, strong UIBs, and the AGN NGC 1068 (bottom): high excitation, no UIBs (Sturm et al 1999b, Colbert et al 1999, Spinoglio et al 1999). Sudden breaks in the SEDs are the result of different aperture sizes at different wavelengths. Local bumps and unusual slopes in the Circinus spectrum (12-20μm and 35μm) may be caused by residual calibration uncertainties.

**Figure 4.** Comparison of the LWS spectrum of Arp 220 and the Galactic star forming region/molecular cloud SgrB2 (Fischer et al 1999, Cox et al 1999). The spectrum of SgrB2 has been divided by a factor of 115 and shifted to the redshift of Arp 220. The main molecular absorption bands are identified, along with the [OI] and [CII] fine structure lines.

**Figure 5a(left):** SWS diagnostic diagram (from Genzel et al 1998). The vertical axis measures the flux ratio of high excitation to low excitation mid-IR emission lines, the horizontal axis measures the strength (i.e. feature to continuum ratio) of the 7.7μm UIB feature. AGN templates are marked as rectangles with crosses, starburst templates as open triangles, UILRGs as filled circles. A simple 'mixing curve' from 0% to 100% AGN is shown with long dashes. 5b(right): ISOCAM diagnostic diagram (from Laurent at el 1999), including 35 CVF spectra from a sample of galaxies of different characteristics. The vertical axis measures the ratio of 14.5 to 5.4μm continuum, the horizontal axis measures the strength of the UIB feature. Active starbursts are in the upper left, AGNs in the lower left and PDRs in the lower right.

The curves from lower right to upper left denote HII/starburst fractions of 25, 50 and 75%. The two curves from lower left to upper right AGN fractions of 75 and 50%.

**Figure 6** a (left): ISOCAM LW2 map (red) of Centaurus A, superposed on a visible image and contours (blue) of the 20cm radio continuum. The infrared data penetrate the dust lane and reveal the presence of a barred dust spiral centered on the AGN (Mirabel et al. 1999). 6b(right): ISOCAM LW3 map (red contours) of the Antennae galaxies (upper right inset shows the the larger scales and the tails) superposed on a V/I-band HST image (Mirabel et al 1998). The nuclei of NGC 4038 and NGC 4039 are top and bottom right, respectively. The interaction region is located below the center of the image and exhibits the strongest mid-IR emission.

**Figure 7.** Left: SWS measurements of the [NeIII]/[NeII] ratio in starburst galaxies (filled rectangles) and nearby starburst templates (asterisks) (from Thornley et al 1999). Top right: [NeIII]/[NeII] line ratios (triangles) and [ArIII]/[ArII] ratios (open rectangles) as a function of Galacto-centric radius for HII regions in our Galaxy (Cox et al 1999). Bottom right: Models of [NeIII]/[NeII] ratio as a function of burst age, for different IMF upper mass cutoffs. For each $M_{up}$ single cluster models for burst durations of 1, 5 and 20 Myr (top to bottom) are shown as continuous lines. The dashed lines are the corresponding 20 Myr models with a cluster mass function.

**Figure 8** a. [CII] observations and PDR models. Ratio of [CII] line intensity to total far-infrared intensity ($Y_{[CII]}$) as a function of ratio of CO 1-0 line intensity to far-infrared intensity ($Y_{CO}$). KAO observations: star forming galaxies (light blue open circles, Stacey et al 1991, Lord et al 1995, 1996), normal galaxies/outer parts of spirals (filled blue circles, Stacey et al 1991, Madden et al 1993), low metallicity galaxies (pink filled rectangles, Poglitsch et al 1995, Madden et al 1997) and Galactic HII regions/PDRs (green crosses, Stacey et al 1991). LWS ISO observations: US Key Project, normal and star forming galaxies (blue, down-pointing open triangles, Malhotra et al 1997, Lord et al 2000), Virgo galaxies with normal and low activity (black open, up-pointing triangles, Smith & Madden 1997, Pierini et al 1999) and ultra-luminous galaxies (filled red circles, Luhman et al 1998, Luhman 1999). Solar metallicity PDR models for clouds of $A_V$=10 and $n_H$=$10^3$ (thick blue line), $10^5$ (dashed red) and $10^7$ cm$^{-3}$ (thin dashed light blue) are from Kaufman et al (1999), with radiation field densities of 10, $10^3$ and $10^5$ times the solar neighborhood field ($G_0$) marked by black, long dashed lines. A model for a $A_V$=3, Z=0.1 cloud is given as green dashed curve.

**Figure** 8b. [CII] and [OI] observations. Dependence of $Y_{[CII]}$ on IRAS 60 to 100 µm flux density ratio (~dust temperature) (from Malhotra et al 1997, 1999a). Right: Line flux ratio of [OI]/[CII] as a function of the IRAS 60 to 100µm ratio (from Malhotra et al 1999a).

**Figure 9.** Average ISOPHOT-S spectra of 20 Seyfert 1 galaxies (top) and 23 Seyfert 2 galaxies (bottom) from the CFA sample (Clavel et al 1998, Schulz and Clavel, priv.comm.). The wavelengths of the main UIB features and of the 9.7µm silicate feature are marked by arrows. UIB features are relatively more prominent in Sey 2s than in Sey 1s but that is mostly due to the stronger red continuum in Sey 1s. Silicate feature emission and absorption are weak.

**Figure 10.** Fraction of sources as a function of ratio of 7.7μm UIB luminosity to 60+100μm-band IRAS luminosity for ULIRGs (filled circles and thin continuous line), starbursts (open circles and thick line), Seyfert 2s (open rectangles and thick dashed) and Seyfert 1s (filled rectangles and thin long dashes) (data from Genzel et al 1998, Rigopoulou et al 1999, Clavel et al 1998, Schulz and Clavel, priv.comm.)

**Figure 11.** UV spectral energy distributions of three Seyfert galaxies re-constrcuted from infrared/optical NLR line ratios (from Alexander et al 1999, 2000, Moorwood et al 1996). The best model is indicated by a thick continuous line and the shaded region marks the 99% (90% for Circinus) confidence zone. In the cases of NGC4151 and NGC 1068 possible composite models with an absorbed bump are shown also.

**Figure 12.** Mid-IR SEDs of ULIRGs. Left: Average spectrum of 60 ULIRGs with z<0.3 (lower inset), compared to average spectra of starburst templates (upper left) and AGNs (upper right). Individual spectra were shifted to the proper rest-wavelength and normalized to give all sources equal weight. The dotted curves in the upper insets show the impact of 50 magnitudes of extinction on the SED shape (from ISOPHOT-S data of Lutz et al 1998a). Right: UIB strength as a function of IR-luminosity of ULIRGs. In addition to the Lutz et al sample (see also Rigopoulou et al 1999), the figure contains 15ULIRGs with z<0.4 from Tran et al (2000) and two hyperluminous ULIRGs from Taniguchi et al (1997b) and Aussel et al (1998), all observed with the CVF of ISOCAM. In the text it is (arbitrarily) assumed that sources above the dotted line (UIB strength ~1) are dominated by star formation, and those below the line by AGNs.

**Figure 13.** Deep mid-infrared surveys. Deep surveys with ISO. Left: ISOCAM 15μm (LW3) contours (gold) on top of WFPC2 image of the HDF (N). 7μm sources (LW2) are marked by green circles (Aussel et al 1999a). Right: LW3 contours on top of R-band image of Abell 2390 (z=0.23) (Altieri et al 1999). The outer regions of the image with less integration time has been down-weighted. With the aid of the foreground cluster's lensing amplification this image reaches to <30μJy for the <z>~0.7 background galaxies. It thus represents the deepest mid-infrared image obtained by ISOCAM.

**Figure 14.** Top left: Summary of integrated 15μm source counts from the different ISO surveys (from Serjeant, priv.comm., Elbaz et al 1999a, normalizing downwards by a factor 1.5 the counts of Serjeant et al 1999), compared to no evolution models (continuous) matching the IRAS counts (the Rush et al 1993 counts were renormalized downwards by a factor of 2). Top right: Differential 15μm counts with the shaded area marking the counts predicted with non-evolution models (from Elbaz et al 2000). The counts are normalized to a Euclidean distribution of non-evolving sources which would have a slope of index -2.5 in such a universe. Data points: A2390 (open stars, Altieri et al 1999, Metcalfe et al 1999), HDF(N) (open circles, Aussel et al 1999a), HDF(S) (filled circles, Oliver et al 1999, Elbaz et al 1999a), Marano (open squares, crosses, stars, Elbaz et al 1999a), Lockman hole (open and filled triangles, Elbaz et al 2000).

**Figure 15.** Cosmic UV to mm, extragalactic background. Open squares give the lower limits from ISOCAM 15μm, ISOPHOT 175μm, and SCUBA 850 μm (Blain et al 1999) sources. The optical-UV points are from Pozetti et al (1998). The COBE FIRAS (grey shaded) and DIRBE 140/240 μm (filled circles) data are from Lagache et al (1999) (from Elbaz et al 2000). Different SEDs are shown and normalized to the 15 μm ISOCAM limit: M82 (continuous), Arp 220 (long dashes). An Arp 220-like SED would significantly overproduce the COBE background. With an M82-like background the ISOCAM galaxies would contribute about 30% to the COBE background. The SED of a typical Seyfert 2 galaxy (and matching the COBE background) is shown dash-dotted. The hatched upper limits in the mid-IR are derived from the lack of attenuation of high energy γ-rays (Hauser et al 1998) (from Elbaz et al 2000).

**Figure 16.** Deep far-infrared surveys. Left: ISOPHOT 90μm (top) and 175μm (bottom) maps of a 44'x44' field LHNW in the Lockman hole (Kawara et al 1998). Right: Integrated 175μm counts from the FIRBACK (Marano and ELAIS fields) (diamonds: Puget al 1999, filled circles: Dole et al 1999) and Lockman field (green asterisk, Kawara et al 1998). The blue line is the model 'E' of Guideroni et al (1998) that includes a strong ULIRG component at high z. The red dotted curve is model 'A' of Guiderdoni et al (1998) that includes strong source evolution but does not include ULIRGs (from Dole et al 1999).

**Figure 17.** Cosmic star formation rate (per unit comoving volume, h=0.6, $q_0$=0.5) as a function of redshift (the 'Madau' plot, Madau et al 1996). The black symbols (with error bars) denote the star formation history deduced from (non-extinction corrected)

UV data (Steidel et al 1999 and references therein). Upward pointing dotted green arrows with open boxes mark where these points move when a reddening correction is applied. The green, four arrow symbol is derived from (non-extinction corrected) H$\alpha$ NICMOS observations (Yan et al 1999). The red, three arrow symbol denotes the lower limit to dusty star formation obtained from SCUBA observations of HDF (N) (Hughes et al 1998). The continuous line marks the total star formation rate deduced from the COBE background and an 'inversion' with a starburst SED (Lagache et al 1999b). The filled hatched blue and yellow boxes denote star formation rates deduced from ISOCAM (CFRS field, Flores et al 1999b) and ISOPHOT-FIRBACK (Puget et al 1999, Dole et al 1999). The light blue, dashed curve is model 'A' (no ULIRGs) and the red dotted curve model 'E' (with ULIRGs) of Guiderdoni et al (1998).